\documentclass[twocolumn]{aastex631}

\usepackage{graphicx}   
\usepackage{amsmath}    

\newcommand{\oversim}[2]{\protect{\mbox{\lower0.5ex\vbox{%
  \baselineskip=0pt\lineskip=0.2ex
  \ialign{$\mathsurround=0pt #1\hfil##\hfil$\crcr#2\crcr\sim\crcr}}}}}

\shorttitle{The pr\'ah}
\shortauthors{Kroupa et al.}

\graphicspath{{./}{figures/}}

\begin{document}

\title{Open star clusters and their asymmetrical tidal tails}

\correspondingauthor{Pavel Kroupa}
\email{pkroupa@uni-bonn.de}

\author[0000-0002-7301-33777]{Pavel Kroupa}
\affiliation{Helmholtz-Institut f\"ur Strahlen- und Kernphysik \\
Universit\"at Bonn, Nussallee 14-16 \\
53115 Bonn, Germany}
\affiliation{Astronomical Institute\\
Charles University, V Holesovickach 2\\
  18000 Praha, Czech Republic}

\author{Jan Pflamm-Altenburg}
\affiliation{Helmholtz-Institut f\"ur Strahlen- und Kernphysik \\
Universit\"at Bonn, Nussallee 14-16 \\
53115 Bonn, Germany}

\author[0009-0008-8138-3790]{Sergij Mazurenko}
\affiliation{Universit\"at Bonn\\
Regina-Pacis-Weg 3\\
53113 Bonn, Germany}

\author{Wenjie Wu}
\affiliation{Helmholtz-Institut f\"ur Strahlen- und Kernphysik \\
Universit\"at Bonn, Nussallee 14-16 \\
53115 Bonn, Germany}

\author{Ingo Thies}
\affiliation{Helmholtz-Institut f\"ur Strahlen- und Kernphysik \\
Universit\"at Bonn, Nussallee 14-16 \\
53115 Bonn, Germany}

\author[0000-0002-8672-3300]{Vikrant Jadhav}
\affiliation{Helmholtz-Institut f\"ur Strahlen- und Kernphysik \\
Universit\"at Bonn, Nussallee 14-16 \\
53115 Bonn, Germany}

\author[0000-0002-1251-9905]{Tereza Jerabkova}
\affiliation{European Southern Observatory\\
 Karl-Schwarzschild-Strasse 2\\
  85748 Garching, Germany}

\begin{abstract}
  Stars that evaporate from their star cluster by the energy
  equipartition process end up either in a leading or a trailing tidal
  tail.  In Newtonian gravitation and for open star clusters in the
  Solar vicinity, the tidal threshold, or pr\'ah, for escape is
  symmetrical, such that the leading and trailing tails are equally
  populated.  The data by six independent teams that applied the
  convergent point method to map out the tidal tails of four open
  clusters (the Hyades, the Praesepe, Coma Berenices and COIN-Gaia13)
  using Gaia DR2 and DR3 are here applied to test for the expected
  symmetry. All tidal tails contain more stars in the leading
  tail. The combined confidence amounts to an $8\,\sigma$
  falsification of the pr\'ah symmetry. The same test using Milgromian
  dynamics leads to consistency with the data. More effort needs to be
  exerted on this matter, but the data indicate with high confidence
  that the tidal pr\'ah of an open star cluster is asymmetrical with
  the corresponding confidence that Newtonian gravitation is
  falsified. Open star clusters depopulate more rapidly in Milgromian
  than in Newtonian dynamics and the COIN-Gaia13 cluster is here found
  to be nearly completely dissolved. In view of these results,
    the wide-binary star test and the Keplerian Galactic rotation
    curve finding are briefly discussed.
\end{abstract}

\keywords{
Open star clusters(1160) --- Star clusters(1567) --- Tidal tails(1701) --- Gravitation(661)
--- Newtonian gravitation(1110) --- Modified Newtonian dynamics(1069)}

\section{Introduction} \label{sec:intro}

The stars in an open cluster with initial mass $M_{\rm oc, 0}$ orbit
chaotically within it with the many weak gravitational encounters
leading to an on-going redistribution of kinetic energy amongst them
as the cluster evolves towards energy equipartition which cannot be
reached. As a consequence of this two-body-relaxational process there
is a near-constant rate of loss of stars across the tidal threshold,
or pr\'ah\footnote{We adopt this term for the tidal threshold in
  connection with Milgromian dynamics following \cite{Kroupa+2022}
  where an explanation of its meaning related to the foundation of
  Prague ``on the threshold to a mystical world'' is provided.},
$\dot{M}_{\rm oc}\propto M_{\rm oc, 0}$ (eq.~12 in \citealt{BM03}).
By virtue of the stars leaking out of their cluster having nearly the
same velocity as the cluster, they are on very similar Galactocentric
orbits thus either drifting ahead or behind the cluster.
\cite{Capuzzo-Dolcetta+2005}, followed by \cite{Montuori+2007,
  Montuori+2008}, were the first to show how the observed tails of
some star clusters in the Galaxy take their shape and how and why the
(observed) under- and over-densities are theoretically found
and explained. Their work was later confirmed and, also, interpreted
in terms of epicyclic orbits \citep{Kuepper+08, Just+2009}. Tidal
tails grow constantly and uniformly in length for open star clusters
on almost circular Galactocentric orbits (fig.~7 in
\citealt{Kuepper+10}).  Owing to the linearity of Newtonian
gravitation and near the Solar circle and at larger Galactocentric
distances, for such star clusters the leading and trailing tails
contain, within Poissonian fluctuations, the same number of stars. The
symmetry of the tail populations has been quantified by
\cite{PflammA+2023}, who show the evaporation process to be stochastic
and describable as a Bernoulli process. This allows quantifying the
number of stars in the leading, $n_{\rm l}$, and in the trailing tail,
$n_{\rm t}$, and also how likely a certain degree of asymmetry is,
given a total number of detected tail stars, $n=n_{\rm l}+n_{\rm t}$.

The notion to test whether Newtonian gravitation is valid on the
scales of open star clusters by using $n_{\rm l}$ and $n_{\rm t}$ was
introduced in \cite{Kroupa+2022} based on the new compact convergent
point (CCP) method developed by \cite{Jerabkova+2021} to map out the
full extent of cluster tidal tals. These latter authors quantified for
the first time the full length of the tidal tails of the Hyades,
Praesepe and Coma Berenice, and later also of NGC$\,$752
\citep{Boffin+22}. By applying Milgromian and Newtonian gravitational
models to these data, \cite{Kroupa+2022} showed that while the
full tails of the Praesepe, Coma Berenices and NGC$\,$752 are
consistent with Newtonian symmetry, the data are also consistent with
Milgromian gravitation. The full-length tails of the Hyades cluster,
however, are inconsistent with the Newtonian symmetry at more than
$5\,\sigma$ confidence (see also \citealt{PflammA+2023}), while being
consistent with Milgromian gravitation.  The Galactic bar does
  not influence the evolution of star clusters orbiting at
  Galactocentric distances larger than $\approx 4\,$kpc
  \citep{RossiHurley2015}. \cite{Thomas+2023} subsequently showed
that the Milky Way's non-axisymmetric bar potential cannot lead to the
observed asymmetry.  The significant Hyades tail asymmetry thus
supports the possibility that the pr\'ah is Milgrom-asymmetric,
namely, that more stars escape per unit time on the Galaxy-near
  side of the open cluster than on the far side.

If this were to be affirmed with additional data then we would be
forced to discard Newtonian gravitation with corresponding
implications for all of galactic, extragalactic and cosmological
research.  The tidal tail data clearly need significant improvement to
either confirm or discard the Milgrom-asymmetric pr\'ah.  The aim here
is to test data on tidal tails of open star clusters that have been
obtained with the established convergent point (CP) method and by
research teams operating independently of the Jerabkova et al. and
Kroupa et al. efforts and prior to the publication of
\cite{Kroupa+2022}.  These latter authors made the observation
  (their sec.~2.2) that the tidal-tail-data extracted for the Hyades,
  Praesepe, COIN-Gaia13 and Coma Berenices by six different teams
  using the standard CP method appear to show an asymmetry by the
  leading tail having more stars than the trailing tail. This
  asymmetry based on data extracted using the CP method was not
  quantified though, and instead the new tidal tail data extracted
  using the new CCP method were used.  Here we return to the previous
  CP-based observations that were obtained prior to the invention of
  the CCP method. Thus, while in \cite{Kroupa+2022} extended tidal
  tail data in the distance range 50~to~200~pc ahead and behind the
  three clusters Hyades, Praesepe and Coma Berenices and in the
  distance range 50~to~130~pc ahead and behind the cluster NGC\,752
  were used, here we restrict the distance range from 10\,pc~to~an
  upper distance value for which data is reported by the six different
  teams that used the CP method. The CP method is well known and finds
  candidate ex-cluster members that co-move with the cluster and that
  are still in the vicinity of the cluster (typically to within about
  150\,pc). The more involved new CCP method instead can identify more
  distant candidate ex-cluster-members. But it relies on a model of
  the tidal tail because the CP method breaks down when the linearity
  assumption is violated since the stars increasingly deviate from the
  cluster centre-of-mass velocity: as they drift further from the
  cluster they are accelerated in the Galactic potential. Here we
  ignore the cluster NGC\,752 used in \cite{Kroupa+2022} because this
  cluster does not have tidal tail information that was published
  prior to the invention of the CCP method. The Milgromian models
  reported in \cite{Kroupa+2022} were consistent with the tidal tail
  data of the three clusters Hyades, Praesepe and Coma Berenices, but
  the extended tidal tail data of the Hyades were asymmetric with more
  than $5\,\sigma$ confidence. The tidal tails of the two other
  clusters did not indicate a strong asymmetry. Here we use the same
  three clusters but rely on the less-extended tidal tail data
  extracted prior to the invention of the CCP method using the more
  robust CP method and we add the cluster COIN-Gaia13 for which such
  data also exist. These data on the tidal tails closer to the open
  clusters are more sensitive to the potential of the cluster plus
  Galaxy combination since they assess the more recent escape of stars
  through the pr\'ah. 

The here-used data and the analysis as to their possible tidal pr\'ah
asymmetry are introduced in Sec.~\ref{sec:data} and in
Sec.~\ref{sec:symmetry}, respectively. The Milgromian and Newtonian
models are compared with these data in Sec.~\ref{sec:models} and the
results are presented in Sec.~\ref{sec:results}. Sec.~\ref{sec:concs}
contains the conclusions with a brief discussion of recent advances
concerning the validity of Milgromian and Newtonian dynamics.

\section{The cluster data}
\label{sec:data}

The astrometric quality of the Gaia data release 2 (DR2) allows stars
to be extracted from the surroundings of four nearby open 
clusters that most likely originated from the respective clusters. The
standard CP method (see \citealt{Jerabkova+2021} and references
therein) was applied by six different teams on the Hyades
\citep{Roeser+19, MeingastAlves19}, the Praesepe \citep{RS19},
COIN-Gaia13 (\citealt{Bai+22}, using Gaia EDR3) and Coma-Berenices
\citep{Fuernkranz+19, Tang+19} open clusters. This method picks-up
likely cluster ex-members based on their very similar space motion as
that of the cluster, and can therefore only map out the ex-members in
the vicinity of the cluster to a maximal distance $d_{\rm max}$ beyond
which the CP approach breaks down. With the new CCP method,
\cite{Jerabkova+2021} introduce a phase-space transformation allowing
also the ex-members in the full-length tidal tails to be found. As
reasoned in Sec.~\ref{sec:intro}, here we resort only to the previous
work based on the CP method, as we aim to be as conservative as
possible in testing gravitational theory on the cluster pr\'ah.

The position and velocity data of the clusters are from the respective
publications, as collated in Table~\ref{tab:clusters}.  The
Galactocentric Cartesian coordinate system $X, Y, Z$ has $X$ including
the Galactic Centre and pointing from the anchor-point towards the
Galactic centre, the explicit definition depending on various authors'
usage with the $X$-anchor-point sometimes being the Sun, the position
of the cluster or local standard of rest (see also
Sec.~\ref{sec:models}). Galactic rotation is in the positive $Y$
direction and the Galactic North pole in the positive $Z$ direction.

\begin{deluxetable*}{ccccccccc}
  \tablenum{1} \tablecaption{Data of the present-day star clusters and
    tidal tails \label{tab:clusters}} \tablewidth{0pt} \tablehead{
    \colhead{Cluster} & \colhead{$T$} & \colhead{$\bar{T}$} &
    \colhead{$b$} & \colhead{$X, Y, Z$} &\colhead{$V_X, V_Y, V_Z$} &
    \colhead{$n_{\rm l}$} & \colhead{$n_{\rm t}$} &$d_{\rm max}$\\
    & \colhead{(Myr)} & \colhead{(Myr)} & \colhead{(pc)}
    &\colhead{(pc)} &\colhead{(pc/Myr)} &\colhead{} & \colhead{}
    &(pc)} \decimalcolnumbers \startdata
  Hyades\citep{Roeser+19} &580--720 & 650 & 3.1 &$-8344.44, 0.06, 10.22$ &$-32.01, 212.37, 6.13$ &234 &184 &175\\
  Hyades\citep{MeingastAlves19} & `` & `` & ``   & ``  & `` &40 &38 &120\\
  Praesepe\citep{RS19} & 708--832 & 770&
  3.7&$-8441.57, -68.90, 127.03$ &$-32.56, 216.53,
  -2.74$ &214 &124 &210 \\
  COIN-Gaia13\citep{Bai+22} & 150--350  &250& 3.4&$-8621.0, 109.3, 65.3$ &$28.86, 243.83, -3.82$ &222 &77 &210\\
  Coma Ber\citep{Fuernkranz+19} & 700--800 &750&2.7&$-8306.71, -5.91, 112.44$ &$9.12, 236.90, 6.43$ &44 &21 &35\\
  Coma Ber\citep{Tang+19} & ``  & ``     & `` & `` & ``   &8 &5 &35\\
  \enddata \tablecomments{Except for Gaia-COIN13, the ages ($T$) and
    Galactocentric position and velocities stem from table~1 in
    \cite{Kroupa+2022}. $\bar{T}$ is the mean of the given ages.  The
    Plummer parameters $b$ are from \cite{PflammA+2023} for the
    Hyades, Praesepe and Coma Berenices. The derivation of the Plummer
    parameter of COIN-Gaia-13 is described in Sect.~\ref{sec:data}
    (the corresponding half-mass radii being
      $r_{\rm h}\approx 1.305\,b$, e.g. \citealt{Kroupa08}).  The
    number of reported candidate stars in the leading and trailing
    tail of each cluster, $n_{\rm l}, n_{\rm t}$, respectively (see
    Fig.~\ref{fig:clusters}), stem from the respective authors
    applying the standard CP method. The maximum extent of the tidal
    tail as obtained by these authors is $d_{\rm max}$. }
\end{deluxetable*}

\begin{figure*}[ht!]
  \plotone{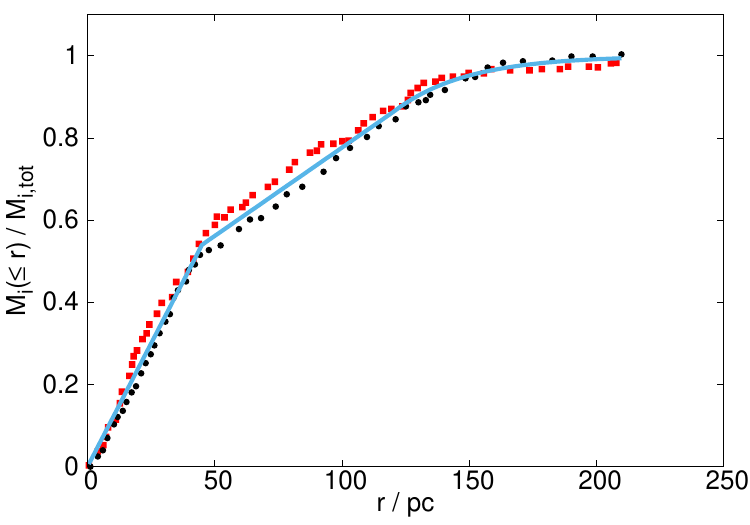}
  \caption{COIN-Gaia13: Radial relative cumulative mass distribution,
    $M_{\rm i}(\le r)$, of stars with masses $m_i<1\, M_\odot$
  (squares) and $m_i \ge 1\,M_\odot$ (circles). The total mass in all
  i~stars is $M_{\rm i, tot}$. The solid line shows the interpolation
  function in Eq.~\ref{eq:coin-gaia-13-cum-mass}.
  \label{fig:coin-gaia-13-cum-mass-fit}}
\end{figure*}

The cluster COIN-Gaia13 needs special attention: The data are obtained
from \cite{Bai+22} who provide a total number of stars of~478 and a
total mass of~$439\,M_\odot$. Based on the local Oort constants they
calculated a tidal radius of 11\,pc. However, the stellar sample is
spread up to a distance of 200\,pc from the cluster centre. The
majority of the stellar mass used for the calculation of the tidal
radius lies outside of the central~$11\,$pc radius.  In order to
derive the proper star cluster (tidal) mass and the corresponding half
mass radius we extract\footnote{{\sc WebPlotDigitizer} at
  \url{https://automeris.io}} the data from fig.~9 (left) in
\cite{Bai+22} which shows the relative cumulative mass fraction of the
stellar sample with a mass less than and
above~1\,$M_\odot$. Fig.~\ref{fig:coin-gaia-13-cum-mass-fit} shows the
relative cumulative radial mass distribution, with $r$ being the
three-dimensional distance from the cluster centre, which is here
interpolated by
\begin{equation}
  \label{eq:coin-gaia-13-cum-mass}
  \frac{M(\le r)}{M_\mathrm{i,tot}} =
  \left\{
  \begin{array}{ccc}
    \rho_{1}\,r &  ;     &    r< r_1 \; ,\\
    \rho_{1}\,r_1 + \rho_{2}(r-r_1) &  ;     &    r_1 \le r < r_2   \;
                                               ,\\
    \rho_1r_1+\rho_2(r_2-r_1) \, +
     \\ (1-\rho_1r_1+\rho_2(r_2-r_1)) \, \times\\ \left(1-e^{-\rho_3(r-r_2)}\right) &
                                                                       ;
                         &   r_2 \le r  \; ,\\ 
  \end{array}
  \right.
\end{equation}
with parameters $r_1 = 45\,\rm pc$, $\rho_1=0.012\,\rm pc^{-1}$,
$r_2 = 120\,\rm pc$, $\rho_2=0.0043\,\rm pc^{-1}$ and
$\rho_3=0.035\,\rm pc^{-1}$, and $M_{\rm i, tot}$ being the total mass
of the $n$ stars.

\begin{figure*}[ht!]
  \plotone{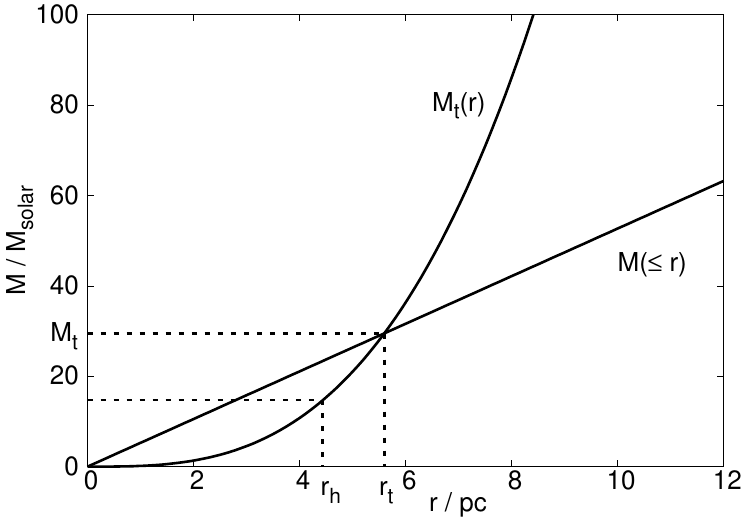}
  \caption{COIN-Gaia13: Observed radial cumulative mass distribution
    of stars in COIN-Gaia13, $M(\le r)$ (as given by the data in
    Fig.~\ref{eq:coin-gaia-13-cum-mass}), and the required 
    tidal mass as a function of the radial distance to the cluster
    centre, $M_\mathrm{t}(r_{\rm t})$ (Eq.~\ref{eq:Mtid}).
    \label{fig:coin-13-tidal_radius}}
\end{figure*}

The required internal tidal mass, $M_\mathrm{t}$,
of a star cluster as a function of the tidal radius, $r_\mathrm{t}$, 
in a disk galaxy with differential rotation is given by \citep{Pinfield+1998}
\begin{equation}
  M_\mathrm{t}(r_{\rm t}) = \frac{2(A-B)^2}{G} \, r_\mathrm{t}^3\,,
\label{eq:Mtid}
\end{equation}
where $A$ ad $B$ are the local Oort's costants.  Here we use
$A=14.5\,\rm km\,s^{-1}\,kpc^{-1}$ and
$B=-13.0\,\rm km\,s^{-1}\,kpc^{-1}$ from \cite{Piskunov+06}. The
intersection of this radial tidal mass function, Eq.~\ref{eq:Mtid},
and the radial cumulative mass distribution
(Fig.~\ref{fig:coin-gaia-13-cum-mass-fit}) determines a tidal mass of
$M_{\rm t}=M_{\rm oc}=29\,M_\odot$ and the tidal radius to be
$r_{\rm t} \approx 5.6\,\rm pc$ (Fig.~\ref{fig:coin-13-tidal_radius},
see e.g. \citealt{Roeser+11} for the method).  It follows that
COIN-Gaia13 has a half mass radius of
$r_\mathrm{h} \approx 4.4\,\rm pc$ and a Plummer parameter
$b\approx 3.4\,\rm pc$. These values supersede those reported by
  \cite{Bai+22}.

\begin{figure*}[ht!]
\plotone{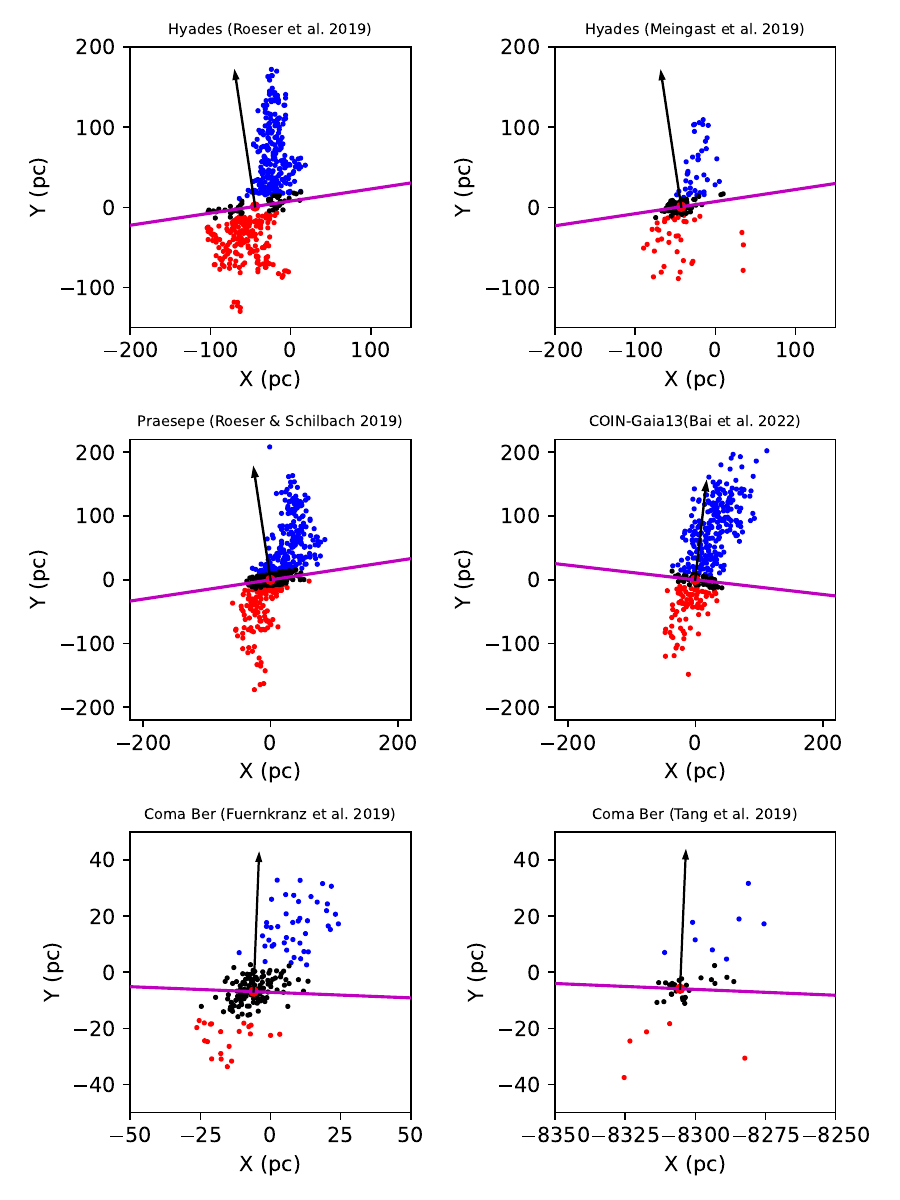}
\caption{The ex-cluster star candidates in the leading tidal tail
  (blue) and trailing tidal tail (red) as obtained by the authors
  referenced at the top of each panel. In all panels, the centre of
  the open cluster is shown as the filled red circle, the
  Galactocentric motion vector of the cluster is shown as the black
  arrow and the line perpendicular to it passing through the cluster
  centre is the solid magenta line. Stars that are further from the
  magenta line than $10\,$pc in the leading tail sum up to $n_{\rm l}$
  and stars in the trailing tail further than $10\,$pc from the
  magenta line sum up to $n_{\rm t}$.  The black stars are closer to
  the magenta line than $10\,$pc and are not used in the present
  analysis. The maximum distance, $d_{\rm max}$
  (Table~\ref{tab:clusters}), probed by the CP method in each case is
  given by the maximum extent of either the blue or red data
  clouds. {\it Upper left panel}: Hyades, as fig.~3 in
  \cite{Roeser+19}. The $X$-anchor-point is the Sun.  {\it Upper right
    panel}: Hyades, as fig.~3 in \cite{MeingastAlves19}.  The
  $X$-anchor-point is the Sun.  {\it Middle left panel}: Praesepe, as
  fig.~2 in \cite{RS19}.  The $X$-anchor-point is the cluster. {\it
    Middle right panel}: COIN-Gaia$\,$13, as fig.~4 in \cite{Bai+22}.
  The $X$-anchor-point is the cluster.  {\it Lower left panel}: Coma
  Berenices, as fig.~2 in \cite{Fuernkranz+19}.  The $X$-anchor-point
  is the Sun.  {\it Lower right panel}: Coma Berenices, as fig.~7 in
  \cite{Tang+19}.  The $X$-anchor-point is the Galactic Centre.}
\label{fig:clusters}
\end{figure*}

Because the open clusters are on near-circular orbits about the
Galactic centre with relatively small excursions above and below the
mid-plane, we analyse the star-counts by projecting the data onto the
$X-Y$ plane, as shown in Fig.~\ref{fig:clusters}. For each of the six
measurements, a line passing through the cluster centre but at right
angle to the respective cluster's velocity vector in the $X-Y$ plane
is constructed and stars that are $10\,$pc ahead and behind this line
in the $X-Y$ plane are counted as being in the leading and trailing
tails, respectively. We use a nominal distance of~10\,pc ahead
  and behind the cluster because this is observationally pragmatic and
  is similar to the tidal radius,
  $r_{\rm tid} \approx \left(0.5\,M_{\rm oc}/M_{\rm
      gal}\right)^{1/3}\,D_\odot \approx 11\,$pc for an open cluster
  weighing $M_{\rm oc}=300\,M_\odot$ at the distance of
  $D_\odot=8300\,$pc from the Galactic centre in a logarithmic
  potential corresponding to a Galactic mass of
  $M_{\rm gal}=0.6\times 10^{11}\,M_\odot$ within $D_\odot$. The
  masses of the clusters in Table~\ref{tab:clusters} are (table~1 in
  \citealt{Kroupa+2022}) $M_{\rm oc}=275\,M_\odot$ (Hyades),
  $311\,M_\odot$ (Praesepe) and $112\,M_\odot$ (Coma Ber), while
  COIN-Gaia13 is derived here to weigh $M_{\rm oc}=29\,M_\odot$.  The
CP method cannot assess ex-clusters stars beyond about $d_{\rm max}$
ahead or behind the clusters (for which the \citealt{Jerabkova+2021}
CCP method is needed) and the star counts ($n_{\rm l}, n_{\rm t}$,
respectively) are thus for the distance range between~$10\,$pc
  and~ $d_{\rm max}$ (as listed in Table~\ref{tab:clusters}) ahead and
  behind the cluster.  This distance rage will be implemented in the
models in Sec.~\ref{sec:models}. From the data in
Table~\ref{tab:clusters} we note that each of the six measurements has
$n_{\rm l}>n_{\rm t}$.

As stars drift away from the open clusters, under- and over- densities
along the star cluster tails form kinematically
\citep{Capuzzo-Dolcetta+2005}. The existence and location of these
under- and over- densities (see figs.~11 and~12 in
\citealt{Capuzzo-Dolcetta+2005}) have been, later, also interpreted in
terms of epicyclic motions relative to the centre of mass of the
cluster \citep{Kuepper+08, Just+2009}.  Overdensities of stars thus
form at regular spacings ahead and behind the cluster, with
\cite{Jerabkova+2021} for the first time reporting evidence for their
existence for an open star cluster. In Newtonian gravitation, the
first epicyclic overdensities have a distance from the centre of mass
position of the open cluster of
$\Delta/{\rm pc} \approx \pm 350\,\,\left(M_{\rm
    oc}/16400\,M_\odot\right)^{1/3}$ (fig.~6 in \citealt{Kuepper+10})
such that $\Delta\approx 64\,$pc for $M_{\rm oc}=100\,M_\odot$ and
$\Delta\approx 51\,$pc for $M_{\rm oc}=50\,M_\odot$, and are thus
within the range $10\,$pc--$d_{\rm, max}$. In Newtonian gravitation,
the overdensities and gaps are spaced symmetrically at equal distances
from the cluster ahead and behind it \citep{Capuzzo-Dolcetta+2005,
  Kuepper+08, Just+2009, Kuepper+10}, but in Milgromian gravitation
the leading overdensity is at a larger distance from the cluster than
the trailing one \citep{Kroupa+2022}. This indicates that the escape
speed is lower towards the leading tail such that stars escape
slightly faster, and thus the relative spacing of the epicyclic
overdensities is a sensitive measure of gravitational theory. For the
purpose here, the number of stars in the leading and trailing tails,
$n_{\rm l}, n_{\rm t}$, respectively, thus contains this
information. If Newtonian gravitation were to be correct, then
$n_{\rm l}\approx n_{\rm t}$ also because the leading and trailing
K\"upper epicyclic overdensities are symmetrically spaced relative to
the cluster's centre of mass position.

\section{Symmetry analysis}
\label{sec:symmetry}

\begin{figure*}[ht!]
\plotone{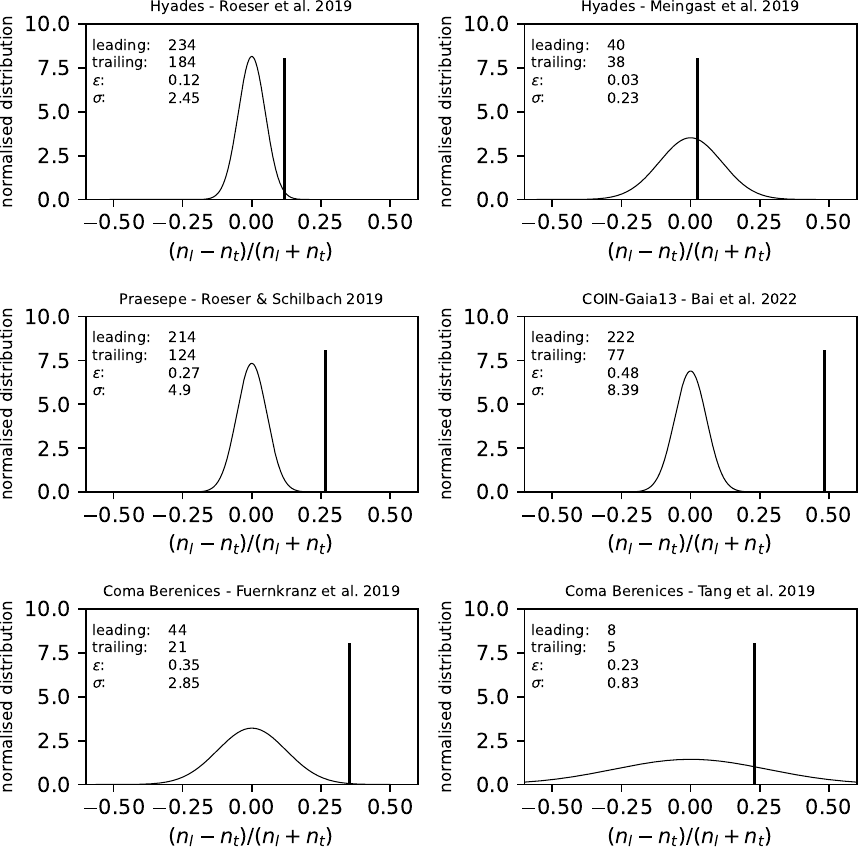}
\caption{Application of the method of assessing the deviation from the
  null hypothesis in terms of $\sigma$ for each of the six
  observations of tidal tails displayed in Fig.~\ref{fig:clusters}.
  The leading and trailing tail numbers for each observed tidal tail
  gauge the probability whether a measurement is one-sided asymmetric
  with $n_{\rm l}>n_{\rm t}$.  See Sec.~\ref{sec:symmetry} for
  details.}
  \label{fig:probs}
\end{figure*}

If the four open clusters are orbiting in a smooth potential and are
thus unperturbed then the stochastic Bernoulli calculation by
\cite{PflammA+2023} can be applied to assess the likelihood of the
observed $n_{\rm l}$ and $n_{\rm t}$ values to occur assuming the null
hypothesis that Newtonian gravitation is valid. Following
\cite{PflammA+2023}, the normalised asymmetry parameter is defined as
\begin{equation}
\epsilon = \left(n_{\rm l}-n_{\rm t}\right)/\left(n_{\rm l}+n_{\rm t}\right) \, .
\label{eq:asymmetry}
\end{equation}
This definition has the advantage that the quantity $\epsilon$ is
symmetrical about $0$.  By the null hypothesis, the expectation value
is $\mu_{\epsilon}=0$ with variance $\sigma_\epsilon =
1/\sqrt{n}$. For each of the six measurements, the asymmetry
significance $\sigma = |\mu_\epsilon - \epsilon|/\sigma_\epsilon$ is
calculated and displayed in Fig.~\ref{fig:probs}. For example, for the
Hyades \citep{Roeser+19} the asymmetry significance is
$\sigma=|0-0.12|\,\sqrt{418}=2.45$.  The data for the Praesepe and for
COIN-Gaia$\,$13 constitute significant deviations from the null
hypothesis, which is rejected with, respectively,~$4.9\, \sigma$ and
$8.39\,\sigma$~confidence.

Is it possible that Newtonian gravitation is valid but that the star
counts and the application of the CP and CCP methods lead to a bias
that creates a number asymmetry of the tidal tails? An interesting
insight is obtained by the different teams reporting, for the same
clusters, rather different numbers of tail members, despite using the
same Gaia data releases (Table~\ref{tab:clusters}).  This indicates
the need to further study the tail-member extraction
algorithms. Nevertheless, while the extracted numbers differ, in all
cases the leading tail has more stars than the trailing
tail. Particularly interesting is that the re-analysis by
\cite{Thomas+2023} of the Hyades using the CCP method but allowing the
model cluster to evolve in non-axisymmetric Galactic potentials
extracts different tail candidates than the original study by
\cite{Jerabkova+2021} who used an axisymmetrical potential. But both
extracted tidal tails have comparable tail lengths and the asymmetry
also remains similar.  The clusters in Table~\ref{tab:clusters} are in
different directions as seen from the Sun, and this suggests that
observational bias will not affect the star counts in the same
manner. If such an effect were there, we might expect a more random
result in terms of which half-tail contains more reported
stars. Again, the finding that all measurements of all clusters
available today indicate the same symmetry breaking suggests that an
observational bias leading to this asymmetry is not likely.

Is it possible that Newtonian gravitation is valid but that the tidal
tail asymmetries are a result of the open star clusters being
perturbed?  According to the data used here, the Praesepe and
COIN-Gaia$\,$13 show a very significant asymmetry
(Fig.~\ref{fig:probs}). This could be due to a perturbation. Indeed,
\cite{Jerabkova+2021} studied the possibility that the Hyades is
heavily perturbed by a recent encounter which lead to the significant
asymmetry of the number of stars in the leading and trailing full-length
tidal tails. A recent encounter with a massive perturber can lead to
an effect comparable to the observed asymmetry of the full-length tidal
tails, but the mass ($\approx 10^7\,M_\odot$) and proximity
($\approx 120\,$pc) appears for this to be unlikely because neither is
a perturbation of the Solar neighbourhood field population known nor
are there any correspondingly massive molecular clouds there, as
pointed out by \cite{Jerabkova+2021}. The analysis here based on the
inner tidal tail data as obtained by the CP method shows the Hyades to
be consistent with the null hypothesis while the Praesepe (a
$4.9\,$sigma deviation) and COIN-Gaia$\,13$ (an $8.39\,$sigma
deviation) are not. Combining the available information: The extended
tidal tails of the Hyades are in $>5\,$sigma tension with the null
hypothesis \citep{Kroupa+2022, PflammA+2023}, and the inner tidal
tails of the Praesepe are in $4.9\,$sigma and of COIN-Gaia$\,13$ in
$>5\,$sigma tension with the null hypothesis. The asymmetry is the
same in all three cases, namely, the leading tail contains
significantly more stars than the trailing tail. The remaining
cluster, Coma Berenices, does not show a highly significant asymmetry
but nevertheless also has $n_{\rm l}>n_{\rm t}$ at the near-$3\,$sigma
confidence level.  While no perturber with a corresponding mass is
evident to be present, the possibility that all of these clusters
suffered an encounter at the same time leading to the same type of
asymmetry appears to not be physically plausible.

\section{Models}
\label{sec:models}

Sec.~\ref{sec:symmetry} documents that there is strong evidence that
Newtonian gravitation may not be universally valid.  In
Sec.~\ref{sec:zexc} Newtonian models of a Hyades-like cluster are
studied to assess if a realistic orbit of an open cluster which
includes passing through the Galactic mid-plane can lead to an
asymmetry of its tidal tails. In Sec.~\ref{sec:MLDcode} Milgromian-
(MOND, \citealt{Mil83, FamMcgaugh, Milgrom14, Merritt20, BanikZhao21})
and Newtonian-dynamics models are studied in order to advance our
knowledge on the tidal tail asymmetry of open star clusters. We refer
to \cite{Kroupa+2022} for a thorough introduction and discussion of
the problem.

\subsection{Tidal-tail asymmetry due to $Z$-excursions?}
\label{sec:zexc}

Three Newtonian simulations (referred to in the following as
  models~1,~2 and~3 with different random number seeds for the stellar
  masses, position and velocity vectors but otherwise identical
  initial parameters) of a Hyades-like star cluster with an initial
  total mass of $1300\,M_\odot$ and an initial Plummer parameter of
  $b=2.3\,$pc (half-mass radius of $3.0\,$pc) are performed to test if
  a realistic Galactocentric orbit that is inclined to the Galactic
  plane might lead to periodic tidal tail asymmetries similar to those
  observed. The direct $N$-body code PETAR \citep{Wang+2020} is
  applied. It is a newly developed high-end code resting partially on
  the developments that lead to the Aarseth-suite of $N$-body models
  such as Nbody6 \citep{Aarseth99, Aarseth10}. PETAR allows precise
  and accurate star--star force calculations and thus the integration
  of stellar orbits by being based on the Barnes-Hut tree method, the
  Hermite integrator and invoking slow-down algorithmic regularisation
  and thus also caters for high initial binary fractions.

  The model clusters are initialised according to the Plummer
  phase-space distribution function \citep{plummer, AHW74}.  Modelling
  open star clusters with the Plummer phase space distribution
  function is motivated by its simplicity (e.g. \citealt{HeggieHut03,
    Kroupa08}) and \cite{Roeser+11} and \cite{RS19} finding it to
  match the observed density profiles of the Hyades and the Praesepe,
  respectively. The Plummer phase-space distribution function is
    also the simplest fully analytical solution of the stationary
    collision-less Boltzmann equation (e.g. \citealt{BT}).

  The computations assume a canonical stellar IMF \citep{Kroupa2001}
  but no initial binary population, as these would significantly slow
  down the calculations without affecting the tidal tail symmetry.  As
  Aarseth's Nbody6 and Nbody7, PETAR incorporates stellar evolution
  using the updated single-stellar evolution/binary-stellar evolution
  (SSE/BSE) algorithms \citep{Hurley+00, Hurley+02, Banerjee+20}.  The
  Milky Way is modelled as an axisymmetric bulge$+$disk$+$dark halo
  potential as given by table~1 (MWPotential2014) in \cite{Bovy2015}.
  All clusters are initially positioned at coordinates corresponding
  to the Hyades, as detailed in Table~\ref{tab:clusters} and the
  integrations of the stellar equations of motion extend up to
  $700\,$Myr.

\begin{figure}
    \centering
    \includegraphics[width=0.5\textwidth]{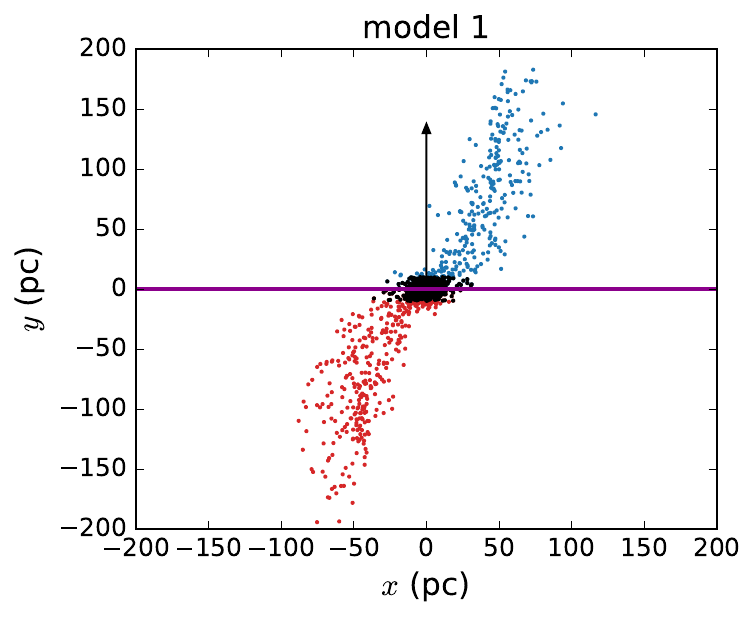}
    \includegraphics[width=0.5\textwidth]{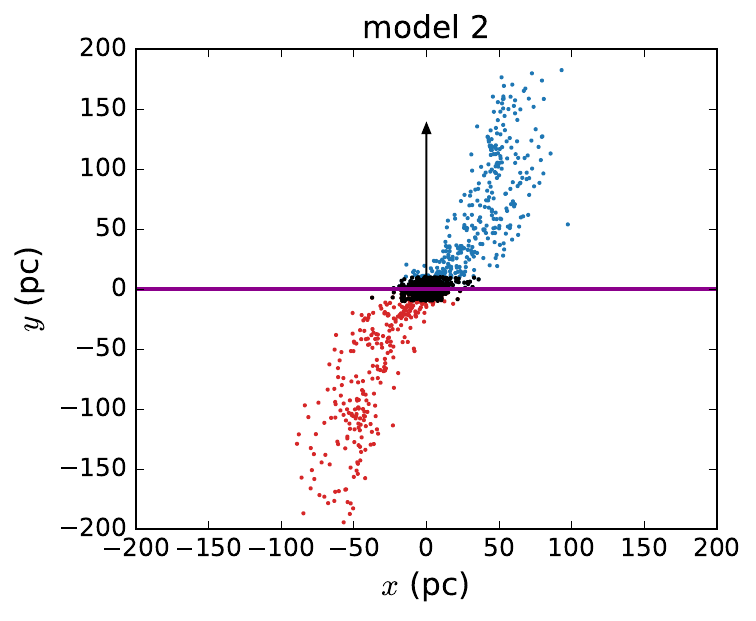}
    \includegraphics[width=0.5\textwidth]{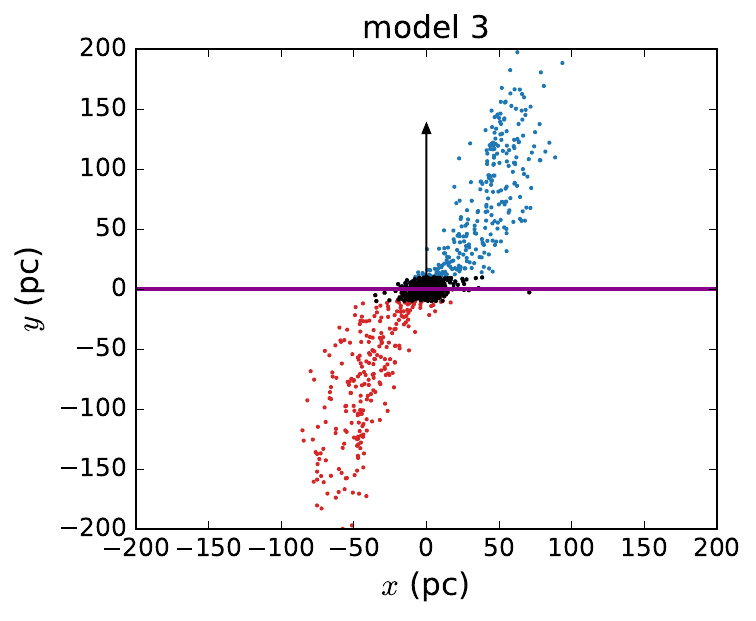}
    \caption{Calculated Newtonian Hyades-like models at $200\,$Myr. Black dots
      represent non-tidal tail stars. Blue dots and red dots denote
      stars in the leading and trailing tails, respectively. Black
      arrows show the Galactocentric motion vector of the cluster and
      the line perpendicular to it passing through the cluster centre
      is the solid magenta line. We show model~1,~2 and~3 from the top to
      the bottom panels. Note that $x=X, y=Y$ (see text).}
    \label{fig:cluster_view}
\end{figure}

  The computations are conducted within a coordinate system that
  combines the Galactocentric system with a translation to the
  cluster's center of mass. To align the model at any snapshot
  with a co-rotating frame, a rotational transformation along the
  $Z$-axis of this system is performed. This adjustment ensures that
  in the resultant $X-Y$ plane, the cluster consistently moves towards
  the positive $Y$-axis, with its center maintained at the
  origin. Tidal tail members are selected as stars in the leading tail
  that have a $Y$-coordinate ranging from $+10$ to $+175\,$pc, and
  those from $-175$ to $-10\,$pc as trailing tail stars (as for the
  Hyades, Table~\ref{tab:clusters}). Fig.~\ref{fig:cluster_view}
  illustrates this approach to selecting tidal tail members at $200\,$Myr.

\begin{figure}
    \centering
    \includegraphics[width=0.5\textwidth]{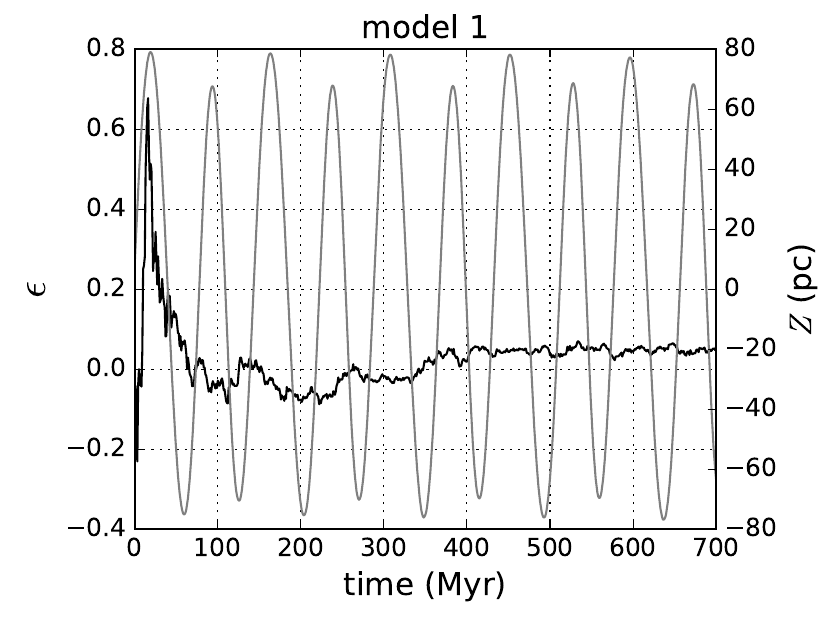}
    \includegraphics[width=0.5\textwidth]{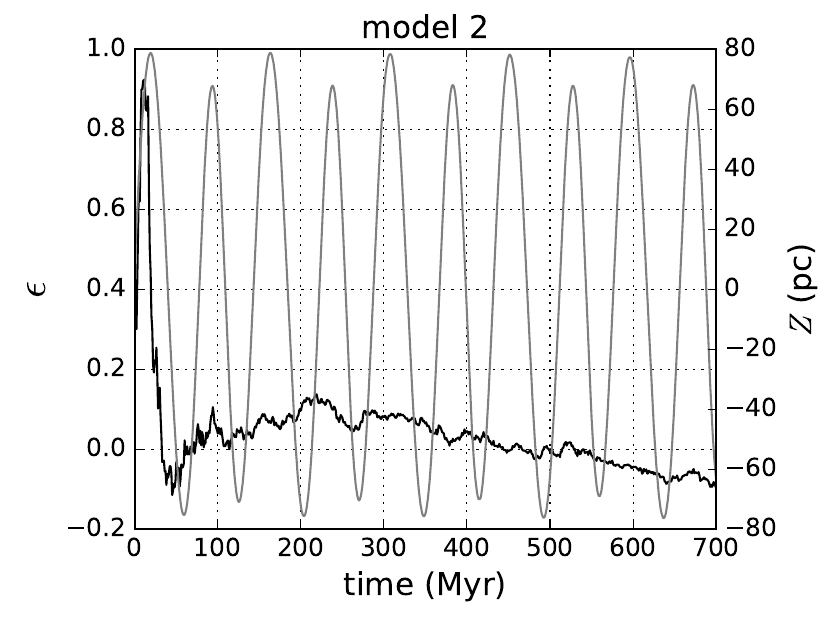}
    \includegraphics[width=0.5\textwidth]{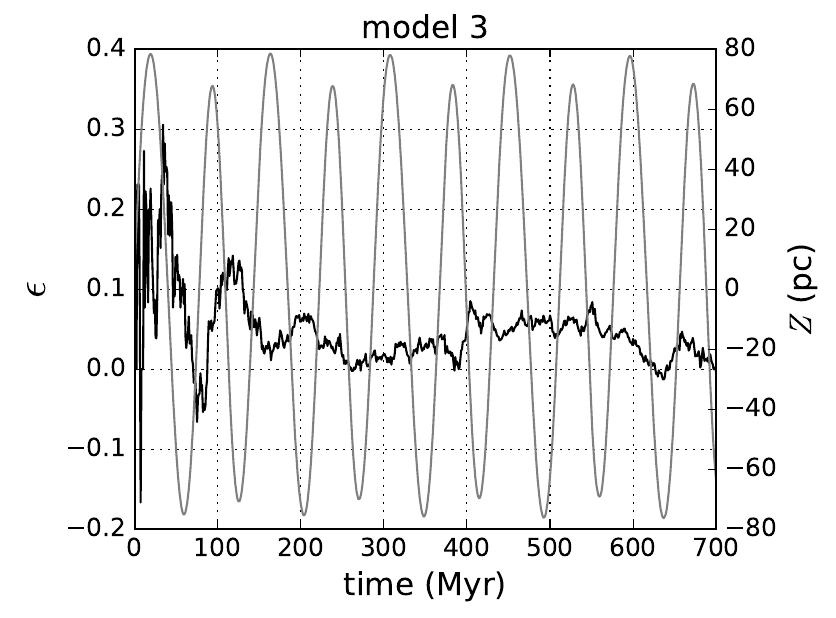}
    \caption{The solid black lines show the time evolution of the
      normalised asymmetry parameter $\epsilon$
      (Eq.~\ref{eq:asymmetry}) for the Newtonian models of
      Sec.~\ref{sec:zexc}. In each panel, the grey line is the
      $Z$-position at each time point. Model~1,~2 and~3 are shown from
      the top to the bottom panels. Note that at the begin of the
      calculation, the tidal tails are not formed, the results before
      $\approx 200\,$Myr therefore not no being reliable.}
    \label{fig:epsilon_plot}
\end{figure}

\begin{figure}
    \centering
    \includegraphics[width=0.5\textwidth]{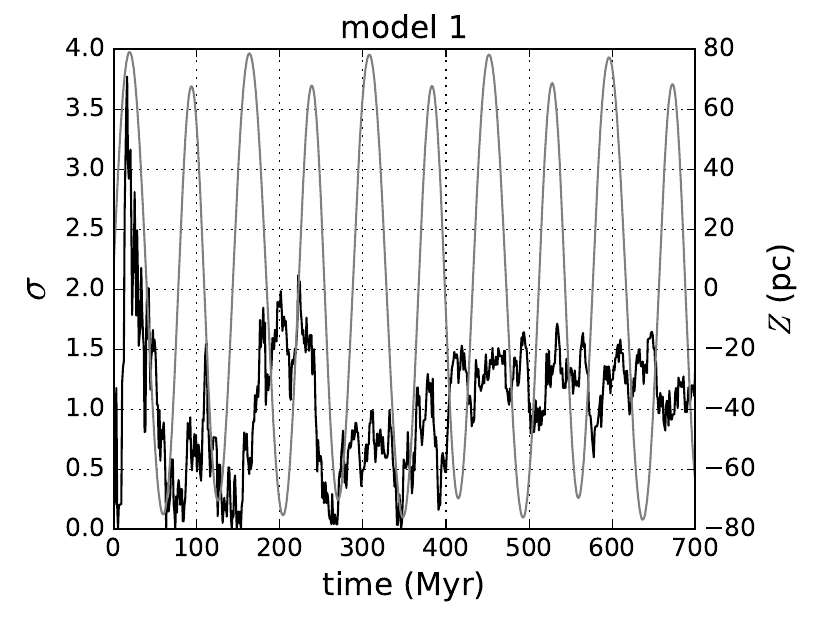}
    \includegraphics[width=0.5\textwidth]{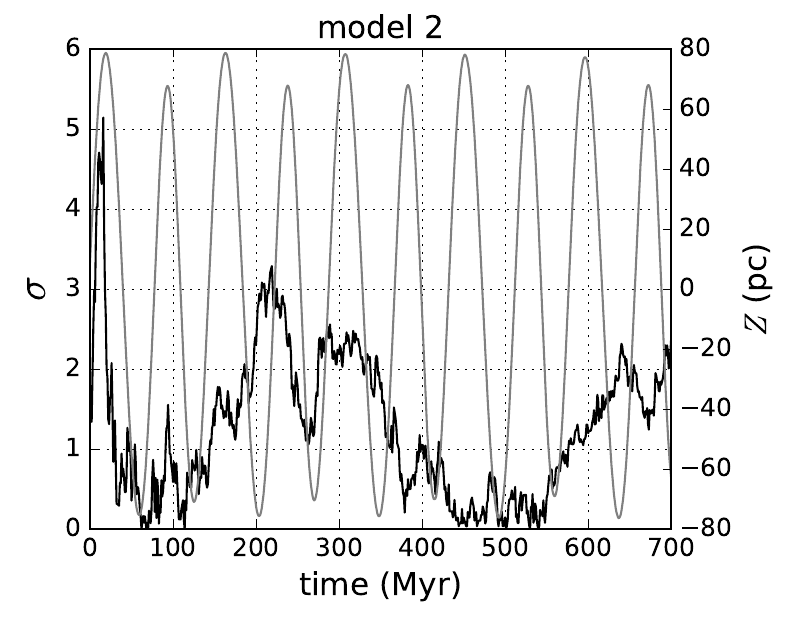}
    \includegraphics[width=0.5\textwidth]{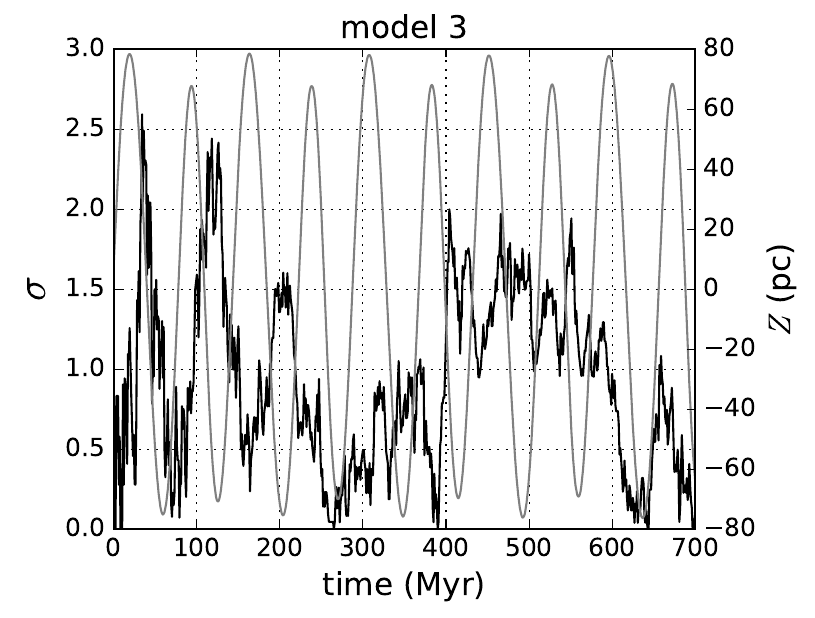}
    \caption{The solid black lines show time evolution of the
      asymmetry significance, $\sigma$ (Sec.~\ref{sec:symmetry}) for
      the Newtonian models of Sec.~\ref{sec:zexc}. The grey line in
      each panel shows the $Z$-position at each time point. Otherwise
      as Fig.~\ref{fig:epsilon_plot}.}
    \label{fig:sigma_plot}
\end{figure}

  At the beginning of the calculations, the tidal tails are not yet
  formed. Consequently, to analyse the asymmetry, it is necessary to
  consider only the snapshots taken after a certain period. This
  determination is based on the scatter plots of stars for each model,
  akin to those illustrated in Fig.~\ref{fig:cluster_view}, but
  capturing various time points. Our analysis indicates that the tidal
  tails are well-developed by $200\,$Myr. Fig.~\ref{fig:epsilon_plot}
  presents the evolution of the normalised asymmetry parameter
  $\epsilon$ (Eq.~\ref{eq:asymmetry}) in conjunction with the
  $Z$-position in the Galactocentric coordinate system. The findings
  demonstrate that excursions in the $Z$-position do not contribute to
  a positive $\epsilon$.  Further analysis on the significance of the
  asymmetry is shown in Fig.~\ref{fig:sigma_plot}. For models~1 and~3,
  the significance consistently remains below $\sigma=2$ once the
  tidal tails have formed. In contrast, model~2 exhibits its highest
  significance, nearly reaching~$\sigma=3$, at around
  $200\,$Myr. However, by $700\,$Myr, the asymmetry parameter
  $\epsilon$ becomes negative. No convincing correlation is evident
  between $\epsilon$, $\sigma$ and $Z$. 

\begin{figure}
    \centering
    \includegraphics[width=0.5\textwidth]{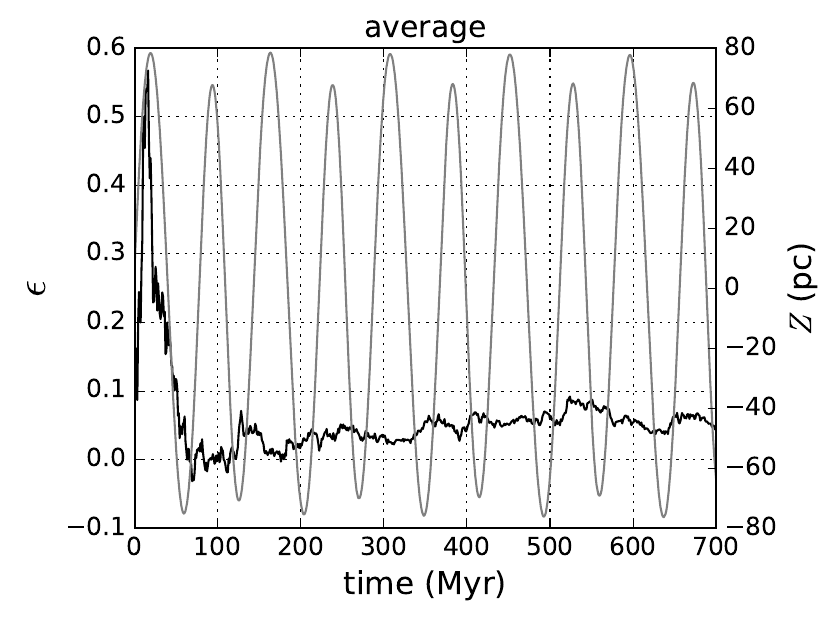}
    \includegraphics[width=0.5\textwidth]{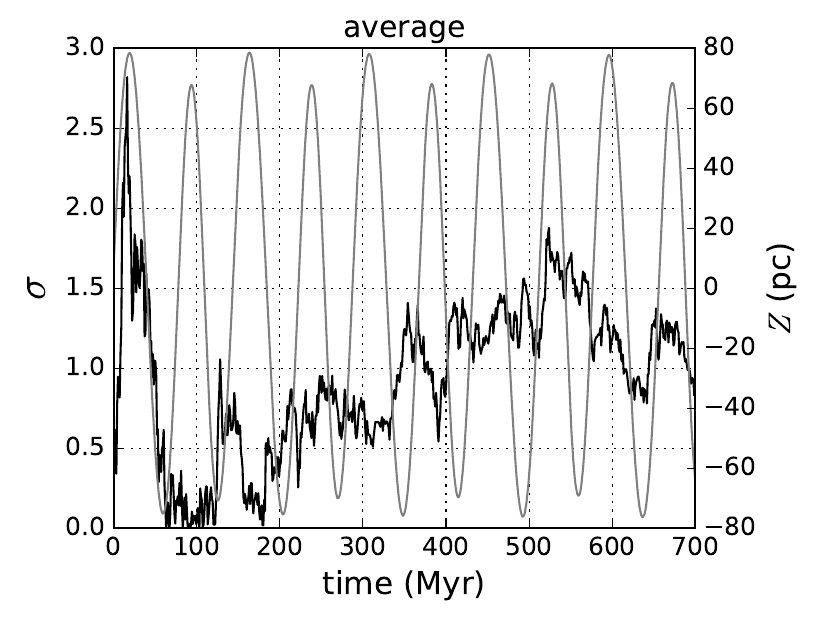}
    \caption{The averaged normalised asymmetry parameter (left, solid
      line, Eq.~\ref{eq:asymmetry}) and averaged asymmetry
      significance (right, solid line, Sec.~\ref{sec:symmetry}) for
      the combined three Newtonian Hyades-like models of
      Sec.~\ref{sec:zexc}. In each panel, the grey line is the
      averaged $Z$-position for the three models 1, 2 and~3. Note that
      there is no significant correlation between $\epsilon$ and
      $\sigma$ with $Z$ and that $\sigma<1.7$ for $T>200\,$Myr in
      agreement with the computational results using the MLD code in
      Newtonian mode of Sec.~\ref{sec:MLDcode}. }
    \label{fig:average_plot}
\end{figure}

  To avoid the stochastic effects stemming from model initialisation,
  the average number of stars in the leading and trailing tails are
  calculated at each snap shot by combining all three
  models. Fig.~\ref{fig:average_plot} displays the resulting asymmetry
  parameter, $\epsilon$, and its significance, $\sigma$. The asymmetry
  parameter indicates a positive asymmetry, meaning the leading tail
  consistently contains more stars than the trailing tail. However,
  the significance of this finding is minimal, remaining below
  $\sigma < 1.7$. A significant correlation with $Z$ is not apparent
  in the model-averaged values of $\epsilon$ and $\sigma$. 

  In conclusion, a realistic orbit of an open cluster oscillating
  about the mid-plane of the Galactic disk therefore does not lead to
  a significant ($\sigma>3$) asymmetry of the tidal tails. Thus
  Milgromian models are considered next in comparison to Newtonian
  models computed with the new MLD $N$-body code.

\subsection{MOND and the MLD code}
\label{sec:MLDcode}

MOND is a non-relativistic theory that generalises Newtonian
gravitation and is non-linear such that the potential around an open
star clusters near the Solar circle is asymmetrical.  In particular,
fig.~3 in \cite{Kroupa+2022} explains why a Milgromian star cluster
looses more stars per unit time into the leading tail than a Newtonian
star cluster.  First of all, in Newtonian gravitation the restoring
force towards the star cluster's centre is equal and opposite on
opposing sides of the cluster's centre such that both tails are fed
equally by evaporating stars.  In contrast, in Milgromian dynamics,
the potential of the cluster on the side towards the Galactic centre
has a reduced restoring force by about 15~per cent (for the point-mass
approximation shown in fig.~3 in \citealt{Kroupa+2022}) towards the
cluster than the backwards side such that more stars can exit it thus
ending up in the leading tail which is fed by stars falling towards
the Galaxy's centre. The fractional reduction of the restoring force
however depends on the mass and extent of the open cluster and on the
particular formulation of MOND used -- see below -- and needs further
theoretical and empirical exploration.  It is also possible that the
escape speeds are very similar but that the pr\'ah on the Galactic
near side has a larger extent than the pr\'ah at the far side such
that more stars can escape towards the Galaxy. Investigations are
on-going as to the details of stellar escape in Milgromian dynamics.
\cite{Kroupa+2022} demonstrated that this pr\'ah asymmetry leads to
the type of asymmetry detected here, namely that the leading tidal
tail of a star cluster has, most of the time, more stars than the
trailing tail. Those models concentrated on the full extent of the
tidal tails in comparison with data extracted using the CCP method,
and the models were idealised by the clusters being set-up on circular
mid-plane orbits in the Galactic disk. Here realistic orbits are
studied considering tidal tail data extracted using the standard CP
method which means only the tidal tails near to the cluster are
assessed (Sec.~\ref{sec:data}). Discussed next, there are three
different formulations of MOND as a gravitational theory that allow
the integration of the equations of motion of particles in a
force-field generated by the matter distribution: AQUAL \citep{BM84},
QUMOND \citep{QUMOND} and MLD \citep{PflammA24}. AQUAL and QUMOND are
field-formulations relating the true (Milgromian) gravitational field,
$\Phi$, to its source, the baryonic matter distribution, $\rho$. MLD
is a particle-based formulation and rests on calculating the particle
forces directly.

Very briefly on AQUAL: consider the following quasilinear elliptic
partial differential equation of second order where the left hand side
is the p-Laplace operator with $u = \Phi_p/a_0$ being the generalised
potential with unit of length,
$ \vec{\nabla} \cdot \left[ |\vec{\nabla} u|^{p-2} \, \vec{\nabla} u
\right] = 4\pi \, G \, \rho/a_0 \, $.  It can be seen that for $p=2$
the standard Poisson equation is obtained with the matter density
$\rho$ sourcing the (Newtonian) potential $\Phi_{p=2}$. For $p=3$,
$\rho$ sources the non-Newtonian potential $\Phi_{p=3}$. In both cases
the negative gradient, $-\vec{\nabla}\Phi_p$, is the acceleration.
The $p=3$ case corresponds to the deep-MOND limit where the equations
of motion are space-time-scale invariant \citep{Milgrom2009}. The full
description includes the transition from the $p=2$ to the $p=3$
Laplace operator when
$|-\vec{\nabla}\Phi| \approx a_0 \approx 3.9\,$pc/Myr$^2$ with
$\Phi=\Phi_{p=2}$ when $|-\vec{\nabla}\Phi| \gg a_0$ and
$\Phi=\Phi_{p=3}$ when $|-\vec{\nabla}\Phi| \ll a_0$. It can be
formulated in terms of a Lagrangian \citep{BM84}. The transition of
non-relativistic dynamics away from Newtonian dynamics to Milgromian
dynamics may be due to the quantum vacuum
\citep{Milgrom99}. Relativistic formulations that encompass these
non-relativistic field equations have been developed (see
\citealt{BanikZhao21} for a review). In \cite{Kroupa+2022} the related
Lagrangian-based quasi-linear formulation of MOND (QUMOND) was applied
to do the simulations because it is computationally more
efficient. QUMOND rests on the idea that the Newtonian potential
generated by $\rho$ can be augmented by a phantom dark matter
potential (which does not consist of real dark matter particles) such
that the combination of both potentials yields the total Milgromian
potential.  The AQUAL and QUMOND formulations of MOND differ when
$|-\vec{\nabla}\Phi_p| \approx a_0$ and this regime is also sensitive to
the transition function between the $p=2$ and the $p=3$ regimes. By
choosing to use QUMOND and a particular transition function,
\cite{Kroupa+2022} explored the general effect of MOND on the escape
of stars across the pr\'ah of their cluster. However, this approach
becomes untenable to model the low-mass open clusters in
Table~\ref{tab:clusters}. Numerical simulations of stellar dynamical
systems by use of field theoretical descriptions require a
sufficiently smooth mass density distribution.  QUMOND simulations
show that the dynamical evolution of already intermediate mass star
clusters ($\approx 5000\,M_\odot$) is effected by the limitation of
the grid resolution and the graininess of the gravitational potential
\citep{Kroupa+2022}.  Therefore, the self-consistent simulation of
star clusters with smaller masses require direct $N$-body methods in a
MONDian context. The above non-linear MONDian field equations have not
yet been discretised such that an $N$-body code has, until now, not
been available to study the dynamical evolution of open star clusters.

In {\it Milgromian Law Dynamics} (MLD), and as a first step towards
such a MOND $N$-body code, Milgrom's law \citep{Mil83} is postulated
to be valid in vectorial form \citep{PflammA24},
\begin{equation}
  \mu(a_{\rm i}/a_0) \, \vec{a_{\rm i}} = \vec{g_{\rm i}}\, ,
  \label{eq:MLD}
\end{equation}
which connects the kinematical acceleration of star~i,
$\vec{a_{\rm i}}$ ($a_{\rm i}=|\vec{a}_{\rm i}|$), to its formal
Newtonian gravitational acceleration, $\vec{g_{\rm i}}$.  The
acceleration $\vec{g_{\rm i}}$ of each particle is obtained from the
sum of the Newtonian gravitational forces from all other particles,
just as in standard Newtonian $N$-body codes (e.g. \citealt{Aarseth99,
  Wang+2020}). The acceleration,
$\vec{a}_{\rm G, i}(\vec{D}_{\rm i})$, acting on particle~i from the
full Galactic potential at the particle's location, $\vec{D}_{\rm i}$,
is added vectorially to $\vec{g_{\rm i}}$. The transition function has
the property $\mu \longrightarrow 1$ for $a_{\rm i}\gg a_0$ and
$\mu \longrightarrow a_{\rm i}/a_0$ for $a_{\rm i}\ll a_0$ such that
Newtonian dynamics is obtained in the former case (e.g. in the
planetary-regime of the Solar system).  The standard transition
function, $\mu(x) = x/\left(1+x^2\right)^{1/2}$ \citep{FamMcgaugh}, is
applied here.  The external field effect (EFE), a unique new physical
phenomenon in MOND and non-existent in Newtonian dynamics
(e.g. \citealt{Haghi+16, Chae+20,Chae+21, Chae+22}), is taken into
account automatically in MLD because the transition function $\mu$ is
on the left-hand side of Eq.~\ref{eq:MLD}.  The MLD code is published
by \cite{PflammA24} where detailed tests are documented and conserved
quantities are derived.

In the MLD code, the standard Hermite scheme used in direct $N$-body
codes \citep{Hut+1995, Makino1991, Kokubo+98, Aarseth99, Aarseth10} is
implemented to integrate the equations of motion of the stellar
particles using the accelerations and jerks in a predictor-corrector
method. The MOND-accelerations, $\vec{a}_{\rm i}$, are obtained by
solving Eq.~\ref{eq:MLD}, and the corresponding jerks are calculated
as the time derivative of the accelerations. In order to avoid the
Newtonisation of the centre of mass of the star cluster in this MOND
formulation and to avoid the handling of rare but
computationally-intensive close encounters, the gravitational $N$-body
force has been softened \citep{Aarseth1963}.  Here, a softening
parameter $\varepsilon=0.1\,$pc is used. This softening does not allow
a realistic assessment of the true evaporation rate which is driven by
the two-body relaxational process. It avoids excessive computational
time but allows the softened particles to self-consistently generate
the Milgromian or Newtonian potentials and thus to map out the
directionality of the relative mass loss from the cluster. In this
sense the MLD code is a first step akin \cite{Aarseth1963} in the
Newtonian case and will be developed further by including
regularisation methods as well as stellar and binary-star evolution
algorithms.  The simulations in pure Newtonian dynamics are performed
with the same MLD code with the same softening by setting the
threshold acceleration to $a_0=3.9\times 10^{-20}\,\rm pc/Myr^2$.

To model the open clusters in Table~\ref{tab:clusters} using the MLD
code, first their initial positions in the Galaxy need to be obtained
by backwards integration to then forward-integrate the initialised
cluster to its presently observed position.  Thus the present-day
position of the star cluster centre is calculated backward in time for
a time equal to the mean estimated age $\bar{T}$ (same procedure as
used in \citealt{Kroupa+2022}). At this position a Plummer phase-space
distribution function is set up as in Sec.~\ref{sec:zexc} with an
initial Plummer parameter $b$ as in Table~\ref{tab:clusters} and
containing $N=2000$ particles of equal mass $m_{\rm
  i}=0.5\,M_\odot$. In the case of the simulations of COIN-Gaia13,
$b=3.4\,$pc and the particle number is $N=500$ with a total mass of
$439\,M_\odot$ (Sec.~\ref{sec:data}) in order to reach the dissolved
state of this cluster.

In the next step each particle~i in the star cluster is integrated
forward in time using the MLD code. All particles are kept in the
calculation, and a spherical logarithmic Galactic gravitational
potential,
$|\vec{a}_{\rm G, i} |= V_{\rm c}^2/\left( X_{\rm G, i}^2+Y_{\rm G,
    i}^2+Z_{\rm G, i}^2 \right)^{1/2}$, is used. It corresponds to a
flat rotation curve of $V_{\rm c}=225\,$km/s, with
$X_{\rm G, i}, Y_{\rm G, i}, Z_{\rm G, i}$ being the Galactocentric
Cartesian coordinates of particle~i. The calculation proceeds until
the density centre of the star cluster comes closest to the current
position of the observed star cluster.  The Solar position in this
coordinate systems is
$X_{{\rm G} \odot}=-8300\,{\rm pc}, Y_{\rm G \odot}=0\, {\rm pc},
Z_{\rm G \odot}=27\,$pc.  The position of the density centre of each
model is found by calculating the density-weighted radius based on the
innermost 20~per cent of particles
\citep{CasertanoHut85,HeggieAarseth1992,KAH}. At this point the model
cluster has a very similar position relative to the Sun and its tidal
tails are extracted just as for the observed clusters in
Sec.~\ref{sec:data}.  In order to obtain sufficient statistics of the
tail occupation numbers, for each of the observed star clusters listed
in Table~\ref{tab:clusters} three models with different initial random
number seeds are calculated in MLD and in Newtonian dynamics.

\section{Results}
\label{sec:results}

\subsection{The observed clusters}

\begin{figure*}[ht!]
\plotone{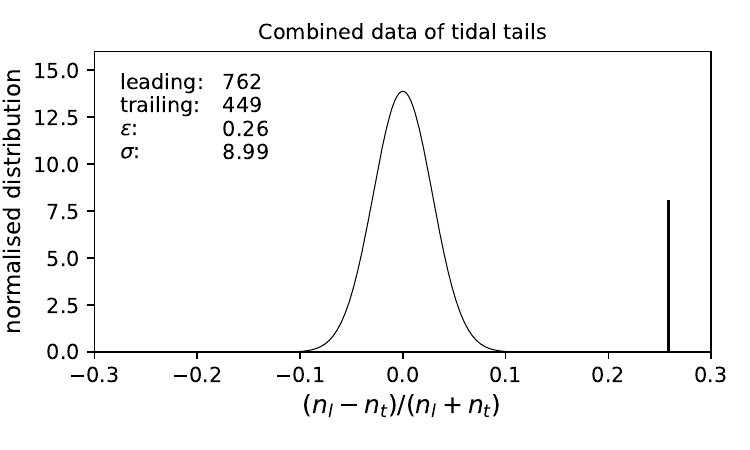}
\caption{As Fig.~\ref{fig:probs}, but here the number of stars in the
  leading and trailing tails are combined from all observed tidal
  tails to assess the probability whether all measurements are
  one-sided asymmetric with $n_{\rm l, sum}>n_{\rm t, sum}$. }
\label{fig:probs_obs_data_combined}
\end{figure*}

In Sec.~\ref{sec:symmetry} the asymmetry-significance, i.e. the
probability of obtaining the number of stars in the leading and
trailing tails, was computed for each observational study of the four
open clusters listed in Table~\ref{tab:clusters}, the results being
documented in Fig.~\ref{fig:probs}.  In order to assess the
probability whether the combined data of the clusters are consistent
with the null hypothesis (symmetrical tails, i.e. Newtonian dynamics
is valid) the observed cluster data are stacked. Since the escape of
stars under the null hypothesis can be very well described as a
stochastic process \citep{PflammA+2023} we can add the leading and
trailing tails in the observed clusters (Fig.~\ref{fig:clusters}),
$n_{\rm l, sum}=\Sigma_{\rm i=1}^6\,n_{\rm l,i}$ and
$n_{\rm t, sum}=\Sigma_{\rm i=1}^6\,n_{\rm ti}$. This combined data
set is extremely significantly discrepant with the null hypothesis
because the available measurements of the tidal tail membership have
significantly more stars in the leading than in the trailing
tail. Based on the observational data, the null hypothesis is
therewith rejected with $8.99\,\sigma$ confidence
(Fig.~\ref{fig:probs_obs_data_combined}).

\subsection{The models}

The above asymmetry-significance is also calculated for each of the
models in order to assess if these confirm the rejection: Are the
tidal tails of the Milgromian models as asymmetric as the observed
ones? And do the Newtonian models confirm the expected symmetry
\citep{PflammA+2023}?

\begin{figure*}[ht!]
  \plotone{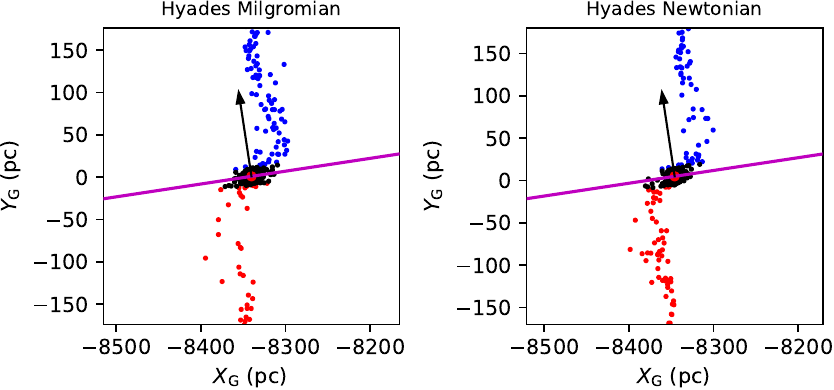}
  \plotone{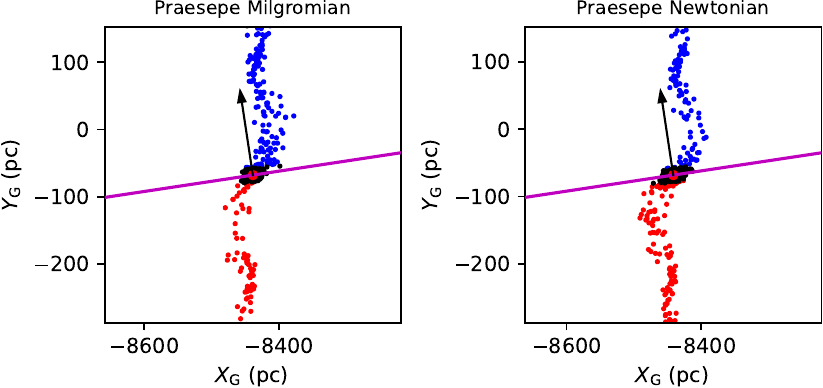}
  \caption{Star cluster models: The three stacked Milgromian (left) and
    three stacked Newtonian (right) models stellar-dynamically evolved
    with the MLD code (Sec.~\ref{sec:models}).  The models are at the
    same location relative to the Sun as the real cluster and
    the symbols and spatial scales are as in
    Fig.~\ref{fig:clusters}  From top to bottom: Hyade, Praesepe.}
  \label{fig:ClusterModels1}
\end{figure*}

\begin{figure*}[ht!]
  \plotone{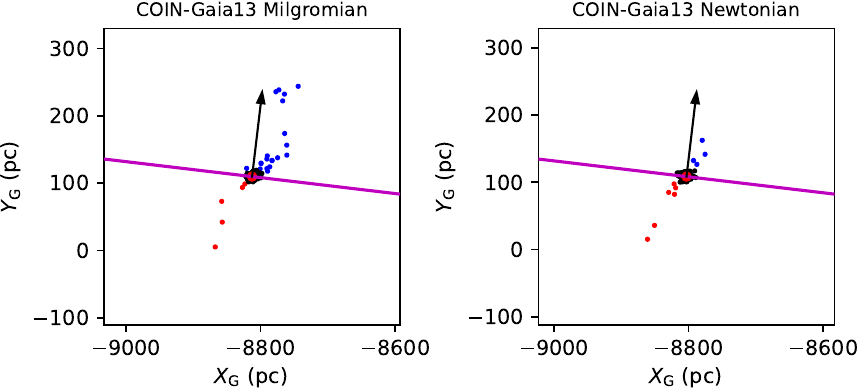}
    \plotone{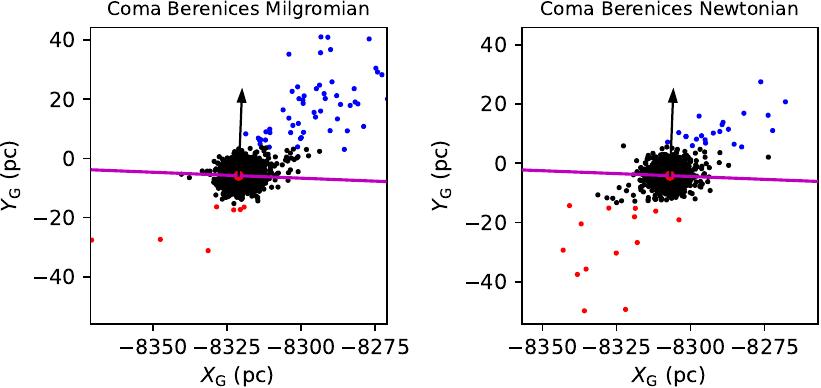}
    \caption{As Fig.~\ref{fig:ClusterModels1} but for COIN-Gaia13,
      Coma Berenices.}
  \label{fig:ClusterModels2}
\end{figure*}

As described in Sec.~\ref{sec:models}, for each open star cluster in
Table~\ref{tab:clusters} three models are computed with the MLD code
in each of the gravitational theories in order to improve the
statistics in the model data.  The final snapshots, when the
respective model is at the position of the observed cluster, are
stacked and shown in Fig.~\ref{fig:ClusterModels1}--\ref{fig:ClusterModels2}.

The distribution of particles on the sky of the stacked models are
shown in Fig.~\ref{fig:onthesky} to illustrate the sky-position, size
and extent of the tidal tails of each of the open star clusters in
Table~\ref{tab:clusters}. It is evident that the Milgromian and
Newtonian models look, at first sight, similar. More subtle
differences can be seen in the case of Coma~Ber which is a close-by
old cluster close to disruption (the Praesepe has a similar age and
contains significantly more stars, Table~{\ref{tab:clusters}).

  \begin{figure*}[ht!]
  \plotone{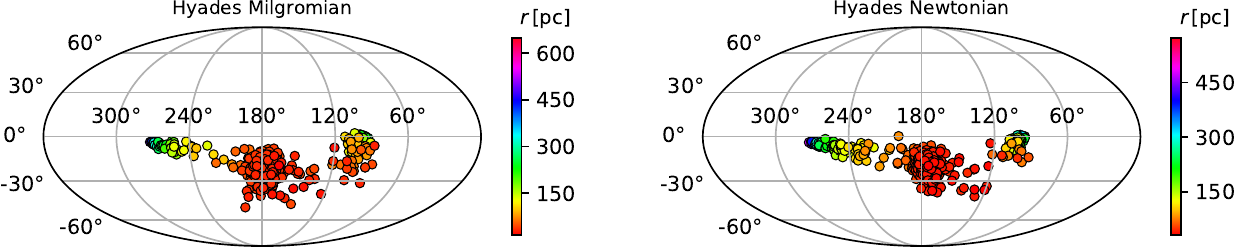}
  \plotone{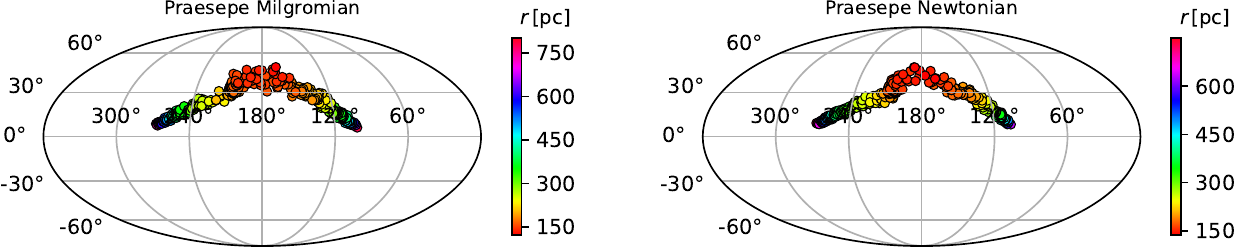}
  \plotone{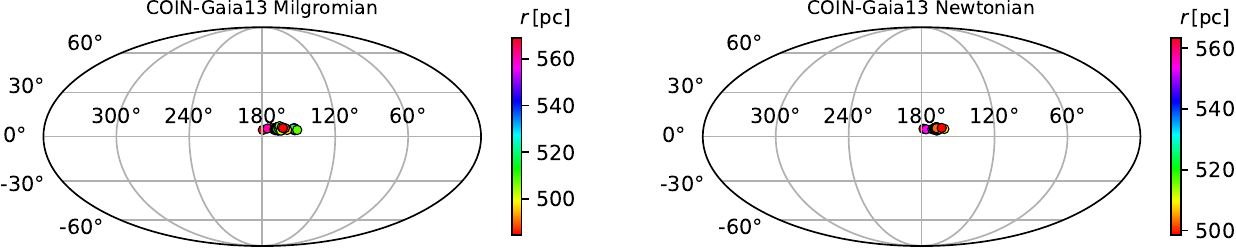}
  \plotone{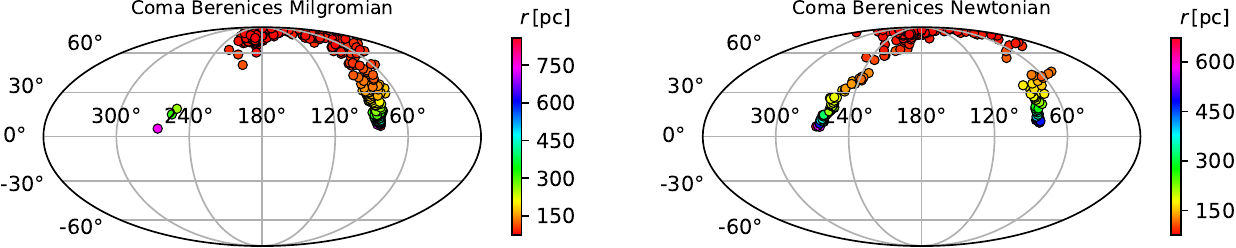}
  \caption{The three stacked Milgromian/Newtonian models of the
    clusters shown in
    Fig.~\ref{fig:ClusterModels1}--\ref{fig:ClusterModels2} are
    presented here in full length as they appear on the model sky in
    Galactic longitude and $l$ and latitude $b$, where
    $l=0^{\rm o}, b=0^{\rm o}$ is the direction towards the Galactic
    centre. The left panels are the Milgromian models and the right
    panels the Newtonian ones. The colour of a stellar particle
    indicates the distance as indicated in the key. As the Galaxy is
    rotating in clockwise direction the leading tails are here on the
    right of the star clusters.  }
  \label{fig:onthesky}
\end{figure*}

The probabilities that the individual stacked models are consistent
with the null hypothesis are evaluated next.  The numbers of particles
and probabilities are shown in Fig.~\ref{fig:indmodelprobs}. It is
already readily apparent, by comparing with the observed clusters
(Fig.~\ref{fig:probs}), that the Milgromian models are indeed highly
inconsistent with symmetric tails, the leading tail always containing
significantly more stellar particles than the trailing one. The
Newtonian models, on the other had, confirm these to be consistent
with symmetrical tidal tails.
\begin{figure*}[ht!]
\plotone{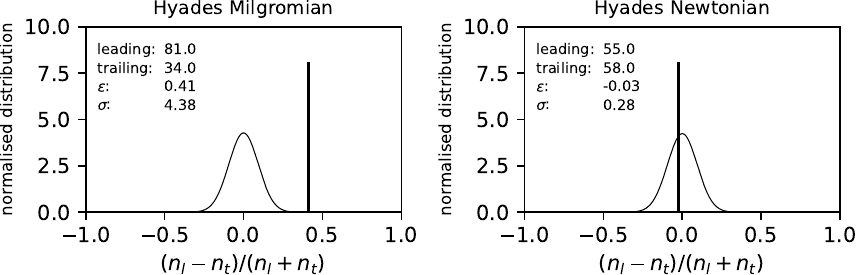}
\plotone{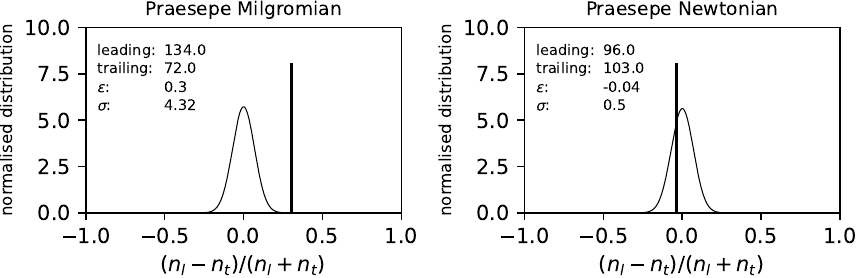}
\plotone{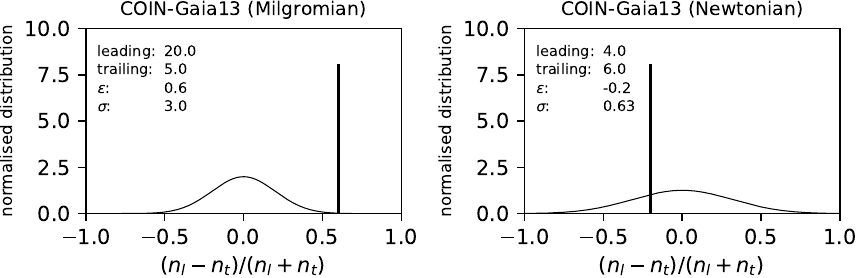}
\plotone{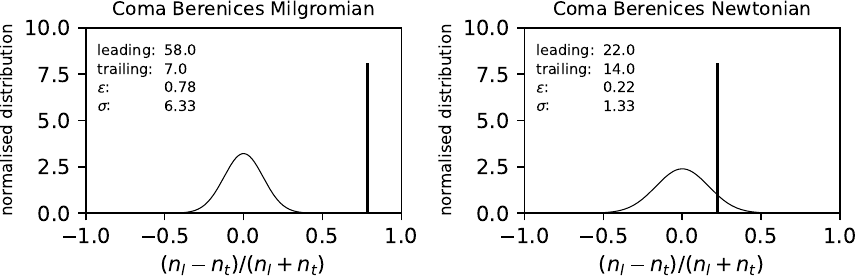}
\caption{As Fig. ~\ref{fig:probs} but here for the models (left/right
  panels for Milgromian/Newtonian). The leading and trailing tail
  numbers of the three stacked models for each case are combined to
  assess the probability whether all measurements are one-sided
  asymmetric with $n_{\rm l}>n_{\rm t}$. All Milgromian models are
  significantly asymmetric ($\sigma\ge 3.0$) with the leading tail
  containing more stellar particles than the trailing tidal tail and
  all Newtonian models are consistent with symmetrical tidal tails
  ($\sigma<1.4$). These Newtonian values are consistent with those
  obtained in Sec.~\ref{sec:zexc} using the PeTaR $N$-body code.}
\label{fig:indmodelprobs}
\end{figure*}
In order to assess the overall probability that Milgromian or
Newtonian models are consistent with the null hypothesis, all
Milgromian and Newtonian models are stacked into one respective
representation, as done for the observed clusters
(Fig.~\ref{fig:probs_obs_data_combined}).  As shown in
Fig.~\ref{fig:combmodelprobs}, the Milgromian models are consistent
with the observed tails by both being significantly dislodged from the
normalised probability distribution, while the Newtonian ones are well
consistent with this distribution. Thus, the Newtonian models computed
with the MLD code confirm that Newtonian tidal tails are symmetrically
occupied, while the Milgromian computations with this same code
confirm the MOND-asymmetry already noted by \cite{Kroupa+2022} on the
basis of a QUMOND code. The combined Milgromian models indeed show a
comparable asymmetry significance ($8.63\,\sigma$,
Fig.~\ref{fig:combmodelprobs}) as the combined observed clusters
($8.99\,\sigma$, Fig.~\ref{fig:probs_obs_data_combined}). Interesting
to note is also that the combined Milgromian models have
$n_{\rm M}=411$ particles in the leading and trailing tails, while the
Newtonian models have $n_{\rm N}=385$ such particles. The difference,
$n_{\rm M}-n_{\rm N}$, corresponds to a three-sigma effect that
suggests Milgromian open clusters to dissolve more rapidly than
equivalent Newtonian ones as discussed in \cite{Kroupa+2022}. This may
be one reason why observed open star clusters are found to be
dissolving more quickly than expected from Newtonian $N$-body models
\citep{Dinnbier+22a}.

\begin{figure*}[ht!]
\plotone{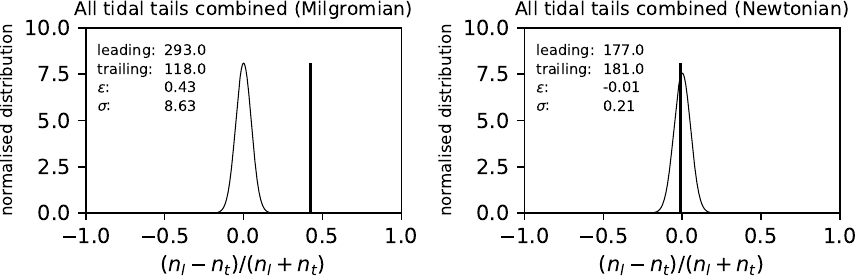}
\caption{Similar to Fig.~\ref{fig:indmodelprobs}. Here the leading and
  trailing tail numbers of all models are combined to assess the
  probability if all measurements are one-sided asymmetric with
  $n_{\rm l}>n_{\rm t}$. The left panel depicts the combined stacked
  Milgromian models and shows a similar and extremely significant
  asymmetry that is very comparable to that observed
  (Fig.~\ref{fig:probs_obs_data_combined}), while the combined stacked
  Newtonian models (right panel) are consistent with the leading and
  trailing tail having a similar number of stars.}
\label{fig:combmodelprobs}
\end{figure*}

The four observed open star clusters that have tidal tail data thus
appear to compellingly indicate Milgromian rather than Newtonian
gravitation to be the valid description of gravitational dynamics on
the pc-scale.

\section{Conclusions}
\label{sec:concs}

The tidal tails of open star clusters near to
  the Sun allow to test gravitational theory. The leading and trailing
  tails have, within statistical uncertainties, the same number of
  stars if Newtonian gravitation is valid. If Milgromian gravitation
  is valid, then the leading tail will have significantly more stars
  than the trailing tail. We use the data from six teams that had
  extracted tidal tail candidate stars for the four nearby open star
  clusters Hyades, Praesepe, COIN-Gaia13 and Coma Berenices using the
  standard CP method that allows to find co-moving ex-cluster member
  stars still in the vicinity of an open star cluster. The available
  data reject Newtonian symmetry with 8.99~$\sigma$ confidence, but
  are well consistent with Milgromian gravitation. 

  The Milky Way's bar potential cannot produce this asymmetry
  \citep{RossiHurley2015,Thomas+2023}, and encounters with massive
  structures also cannot simultaneously account for the similar
  asymmetries observed in open star clusters that are at different
  locations around the Sun. Newtonian simulations performed here show
  that a Hyades-like cluster which periodically oscillates through the
  Galactic disk over 700~Myr never shows a significant asymmetry as a
  consequence of the disk crossings.  While star-cluster simulations
  in Newtonian gravitation cannot explain this asymmetry, simulations
  in Milgromian gravitation naturally produce the observed asymmetry.
  Further tidal-tail data are needed for confirmation and additional
  Newtonian modelling is required including perturbations of the Milky
  Way potential through its spiral arms \citep{Thomas+2023} to sharpen
  these results.

The Milgromian and Newtonian $N$-body computations presented here
support Milgromian open clusters to be dissolving more rapidly than
their Newtonian counterparts, and the present analysis finds the open
cluster COIN-Gaia13 to be nearly completely dissolved.  Open star
clusters near the Galactocentric distance of the Sun are strongly
subject to the EFE because the external field from the Galaxy is
comparable to $a_0$ and the internal acceleration of the open clusters
is much smaller than $a_0$ \citep{Kroupa+2022}.  The results here
  suggest MOND rather than Newtonian dynamics to be relevant for
  understanding the dynamical evolution of open star clusters. But the
  relatively modest available data requires a significant further
  effort on obtaining more tidal tail data in conjunction with
  computer modelling in order to constrain the correct formulation of
  Milgromian dynamics and the transition function.  Additional tidal
  tail data will become available for open star clusters at larger
  distances from the Sun than the currently available clusters. Open
  clusters that are ahead of the Sun in terms of Galactic rotation
  will allow Gaia data to assess their trailing tails with more
  accuracy and precision than the leading tails, while open clusters
  behind the Sun will allow a more accurate and precise mapping of
  their leading arms than their trailing arms. This leads to a bias
  that needs to be catered for in the tail-symmetry analysis.

  Placing the above findings into a broader context, is Milgromian
  dynamics relevant beyond open star clusters?  The Hubble Tension has
  been shown to be resolved by galaxies falling in a Milgromian
  gravitational field to the sides of a Gpc-sized local void we are
  in, a void that is not possible in the standard dark-matter-based
  cosmological model but readily forms in a MOND cosmological model
  \citep{Haslbauer+2020,Mazurenko+2024}.

  Galaxy clusters have been posing some tension in that a factor of
  two in mass appears to be missing but can be accounted for in MOND
  if sterile neutrinos exist (e.g. \citealt{BanikZhao21} and
  references therein) and also if appropriate boundary conditions are
  taken into account for these large structures
  \citep{LopezCorredoira+2022}.  The prominent Bullet cluster of
  galaxies has been used as a default object for the proof of the
  existence of dark matter particles. But the fulfilment of
  hydrostatic equilibrium required for the mass determination of the
  hot gas seems to be problematic \citep{PflammA24}. Problems of
  explaining the Bullet (and the El Gordo) galaxy clusters in the
  standard $\Lambda$CDM cosmological model of structure formation have
  been noted. These galaxy clusters are, however, well understood in a
  Milgromian cosmological model \citep{KraljicSarkar2015,
    Asencio+2021, Asencio+2022}.

It is already well established that Milgromian gravitation correctly
accounts for the properties of elliptical galaxies \citep{Eappen+2022,
  Eappen+2024}, and of disk galaxies (e.g. \citealt{FamMcgaugh,
  BanikZhao21}; specific example: M33, \citealt{Banik+20}; natural
formation of exponential disk galaxies: \citealt{Wittenburg+20};
star-formation properties: \citealt{Nagesh+2023}). The availability of
Gaia DR3 has allowed independent teams to assess the rotation curve of
the Galaxy at Galactocentric distances of~19 to
27~kpc. \cite{Labini+2023, Jiao+2023, Wang+2023} and \cite{Ou+2024}
all find it to be decreasing over this distance range by about
$30\,$km/s being consistent with a Keplerian decline. While a flat
MONDian rotation curve is rejected with $3\,\sigma$ confidence by
\cite{Jiao+2023}, \cite{Ou+2024} stress that globular clusters and
satellite galaxies lead to significantly higher rotation speeds at
distances from~$\approx 15$ to $\approx 200\,$kpc. The Keplerian
fall-off between~$\approx 19$ and $\approx 27\,$kpc is associated with
divergent stellar radial velocity components (fig.~8 in
\citealt{Wang+2023}) which compromises simple solutions of the Jeans
equation and suggests a strongly perturbed outer Galactic disk. The
Keplerian fall-off is also associated with a break of the stellar
surface density at about $17\,$kpc with a steep radial decline to
larger Galactocentric distances (fig.~4 in
\citealt{Labini+2023}). This feature is very similar to such a break
at~$20\,$kpc and similar decline of the disk surface density in
Milgromian models of the Galaxy that involve an encounter with
Andromeda~$\approx 10\,$Gyr ago (fig.~3 in \citealt{Bilek+2018}). Such
Milgromian models of the dynamical history of the Local Group need
more exploration for the exact timing, close-encounter distance and
initial galaxy configurations (e.g. radii and masses of the
pre-encounter galactic disks).  But we already know that they
naturally explain the observed thin/thick disk components, warps of
the Milky Way and of Andromeda disks, the planar satellite galaxy
arrangements around both of them, while also being consistent with the
present-day inclinations of the Galactic and Andromeda stellar and
gaseous disks as well as their relative distance and velocity of
approach \citep{Bilek+2018, Banik+2022}.

Recent work on dwarf spheroidal satellite galaxies
\citep{SafarzadehLoeb2021} reports difficulties in matching their
kinematical data by MOND. This problem remains unsolved in Milgromian
dynamics needing attention, but does not imply that Newtonian
solutions with dark matter exist
(e.g. \citealt{Kroupa1997,McGaughWolf2010}).

On the scale of thousands of~AU, the very-wide-binary-star test has
been shown to falsify Newtonian dynamics with the data being
consistent with Milgromian dynamics \citep{Hernandez2012b,
  Hernandez+2019, Hernandez+2022, Chae2023, Hernandez2023,
  Hernandez+2024, Chae2024}.  Contrary to these results,
\cite{Banik+2024} find that very wide binary stars disprove MOND. This
was rebutted by \cite{HernandezChae2023}, who stress that the
  sub-sample of close wide-binaries need to be shown by the same
  method to comply with Newtonian solutions. This gauging of the
  wide-binary-star test was not demonstrated by \cite{Banik+2024}. 
While the alignment of orbital elements of outer Solar System bodies
are reported to probably be due to MOND \citep{PaucoKlacka2016,
  Pauco2017, BrownMathur2023},  \cite{Vokrouhlicky+2024} find this
  to not be possible, and on the Saturnian distance scale
\cite{Desmond+2024} report problems of matching the planet
  position data with Milgromian dynamics.  Relevant in this work
  is that the field-equation underlying MOND (Sec.~\ref{sec:models})
  needs a discretised analogy for the application to few-body dynamics
  which so-far has not been discovered.  The technical details of the
  above modelling on the scale of thousands of~AU may thus contain
  inconsistencies such that the conclusion reached on the validity or
  non-validity of Milgromian dynamics in this regime remain to be
  questionable \citep{PflammA24}.

  It would appear to be implausible that Milgromian dynamics be valid
  rather than Newtonian dynamics and that dark matter particles that
  are not part of the standard model of particle physics also
  exist. Indeed, apart from the above and apart from the fact that
  dark matter particles have not been found experimentally despite
  40\,yr of search, tests for the existence of dark matter particles
  that had been hoped to be accounting for the apparent non-Newtonian
  phenomena on galaxy scales have been yielding negative results: the
  presence of dark matter halos has been significantly questioned by
  the properties of dwarf galaxies in the Fornax galaxy cluster
  \citep{Asencio+2022}. Applying the Chandrasekhar dynamical friction
  test to various galaxy systems \citep{Kroupa15, Kroupa+2023}, and
  most recently on the Milky-Way/Large-/Small-Magellanic-Cloud triple
  system \citep{OehmKroupa2024}, shows no orbital solutions to be
  possible as the systems merge too quickly to be consistent with
  their observed configuration in phase-space. Solutions without
    dark matter particles but with Milgromian potentials are readily
    obtained though. Noteworthy in this context is that earlier work
    had already shown that the observed rotation curves of disk
    galaxies cannot be reproduced if the theoretically-predicted dark
    matter halos are assumed \citep{McGaugh2005, McGaugh+2007}, and
    elliptical galaxies take too long to assemble in the
    dark-matter-based structure formation models to be consistent with
    their rapid early formation \citep{Eappen+2022}.  Independently of
    the above results concerning Milgromian dynamics, these tests thus
    suggest that dark matter particles do not exist.

    In summary, the overall data situation thus indicates that the
    validity of the universal law of Newtonian gravitation is
    challenged, with Milgromian dynamics apparently accounting for the
    observed celestial dynamics.  Successful structure-formation
    simulations in a Milgromian cosmological model have been published
    (particle-based: \citealt{Katz+13}, and hydrodynamical:
    \citealt{Wittenburg+2023}).  The Milgromian-interpretation of
    dynamics requires significant further attention and the
    implications of these findings for galactic astrophysics and
    cosmology are major if confirmed by further research. Hereby we
    need to keep in mind that different formulations of MOND exist due
    to insufficient knowledge which formulation is correct, and that
    additional differences in the detailed dynamical phenomena arise
    through different transition functions which are also not well
    understood, and that it remains possible that the formulation of
    MOND as is known today may be a simplification of a deeper
    matter--space-time coupling not fully understood at the
    present. This might be the case for example if MOND is related to
    the properties of the quantum vacuum (e.g. \citealt{Milgrom99})
    that may change under different conditions.

    \vspace{5mm}

  We thank an anonymous referee for very helpful comments.  Vikrant
  Jadhav acknowledges support through the Alexander von Humboldt
  Foundation in the form of an AvH Research Fellowship. Wenjie Wu
  acknowledges support through a studentship from the stellar
  populations and dynamics research (SPODYR) group at the University
  of Bonn. We thank the DAAD-Bonn-Prague exchange programme at the
  University of Bonn for support.  Much of this manuscript was written
  at the Department of Astrophysics, Astronomy and Mechanics at
  Aristotle University of Thessaloniki, and PK thanks Padelis
  Papadopoulos and other institute members there for their kind
  hospitality.  This work relies on data obtained by the Gaia
  astrometric space mission. Stacy McGaugh's Triton Station was an
  important source of information on the problem concerning the
  Gaia-DR3-derived Keplerian fall-off of the Galaxy's rotation curve.


\bibliography{refs_tailpap}{}
\bibliographystyle{aasjournal}

\end{document}